\documentclass[11pt]{article}  
 \pdfoutput=1 
\usepackage{tikz}
\usepackage[compat=1.1.0]{tikz-feynman}
\usetikzlibrary{shapes,positioning,automata,backgrounds,calc,er,patterns}
    \usepackage{pgfplots}
\usetikzlibrary{positioning,arrows}
\usetikzlibrary{decorations.pathmorphing}
\usetikzlibrary{decorations.markings}
\usepackage[left=3.cm, right=3cm, top=3cm, bottom=3cm]{geometry}
 \usepackage{tabu}
\usepackage{jheppub}
\usepackage{youngtab}
\usepackage{ytableau}
\usepackage[utf8]{inputenc} 
\usepackage{subcaption}
\usepackage{cancel}
\usepackage{xfrac}
\usepackage{float}
\usepackage{tikz}
\usepackage{titlesec}
\usepackage{fancyhdr}
\usepackage{eufrak}
\usepackage{verbatim}
\usepackage{color, colortbl}
\usepackage[section]{placeins}
\usetikzlibrary{positioning,arrows}
\usetikzlibrary{decorations.pathmorphing}
\usetikzlibrary{decorations.markings}
\usepackage{graphicx,rotating,colortbl}
\usepackage{hyperref,slashed,amsfonts}
\usepackage{color}
\usepackage{amsmath,amssymb,bbold}
\usepackage{mathtools}
\usepackage{multirow}
\usepackage{simplewick}
\usepackage{simpler-wick}

\def\circa#1{\,\raise.3ex\hbox{$#1$\kern-.75em\lower1ex\hbox{$\sim$}}\,}

\usepackage{mathtools}
\usepackage{multicol}
\usepackage{color}
\usepackage{blkarray}
\usepackage{lmodern}
\definecolor{rosso}{cmyk}{0,1,1,0.4}
\definecolor{rossos}{cmyk}{0,1,1,0.55}
\definecolor{rossoc}{cmyk}{0,1,1,0.2}
\definecolor{blu}{cmyk}{1,1,0,0.3}
\definecolor{blus}{cmyk}{1,1,0,0.6}
\definecolor{bluc}{cmyk}{1,1,0,0.1}
\definecolor{verde}{cmyk}{0.92,0,0.59,0.25}
\definecolor{verdec}{cmyk}{0.92,0,0.59,0.15}
\definecolor{verdes}{cmyk}{0.92,0,0.59,0.4}

\newcommand{\svdots}{%
  \vbox{
    \scriptsize \baselineskip 2pt \lineskiplimit 0pt
    \hbox {.}\hbox {.}\hbox {.}\kern-0.75pt
  }%
}

\newcommand{\TeV}{\,{\rm TeV}}

\newcommand{\Tr}{\,{\rm Tr}}

\def\circa#1{\,\raise.3ex\hbox{$#1$\kern-.75em\lower1ex\hbox{$\sim$}}\,}

\newcommand{\beq}{\begin{equation}}
\newcommand{\eeq}{\end{equation}}

\newcommand{\bea}{\begin{eqnarray}}
\newcommand{\eea}{\end{eqnarray}}
\newcommand{\be}{\begin{equation}}
\newcommand{\ee}{\end{equation}}
\font\tenrsfs=rsfs10 at 12pt
\font\sevenrsfs=rsfs7
\font\fiversfs=rsfs5
\newfam\rsfsfam
\textfont\rsfsfam=\tenrsfs
\scriptfont\rsfsfam=\sevenrsfs
\scriptscriptfont\rsfsfam=\fiversfs
\def\mathscr#1{{\fam\rsfsfam\relax#1}}

\newcommand{\SO}{\,{\rm SO}}
\newcommand{\SU}{\,{\rm SU}}
\newcommand{\SP}{\,{\rm Sp}}
\newcommand{\U}{\,{\rm U}}

\newcommand{\ct}{c_\theta}
\newcommand{\st}{s_\theta}
\newcommand{\ctt}{c_{2\theta}}
\newcommand{\stt}{s_{2\theta}}

\newcommand{\gtc}{G_{\text{TC}}}
\newcommand{\ntc}{N_{\text{TC}}}
\newcommand{\mF}{\mathcal F}

\newcommand{\spur}[2]{\psi^{#1}\phantom{}_{#2} }
\newcommand{\spurbar}[2]{\bar{\psi}^{#1 #2}}

\newcommand{\bLgbL}{\bar b_L \gamma^\mu b_L}
\newcommand{\tLgtL}{\bar t_L \gamma^\mu t_L}
\newcommand{\tRgtR}{\bar t_R \gamma^\mu t_R}
\newcommand{\bRgbR}{\bar b_R \gamma^\mu b_R}

\newcommand{\tLgbL}{\bar t_L \gamma^\mu b_L}
\newcommand{\bLgtL}{\bar b_L \gamma^\mu t_L}

\newcommand{\SDS}[2]{(\Sigma^\dagger \overleftrightarrow{D}^\mu\Sigma)_{a_{#1}}\phantom{}^{a_{#2}}}
\newcommand{\DS}[2]{(D^\mu\Sigma)^{a_{#1}a_{#2}}}
\newcommand{\DSd}[2]{(D_\mu\Sigma^\dagger)_{a_{#1}a_{#2}}}

\title{Real and Complex Fundamental Partial Compositeness }
\author{Alessandro Agugliaro,}\author{Francesco Sannino}
 \affiliation{CP$^3$-Origins, University of Southern Denmark, Campusvej 55, Odense M, 5023 Denmark}
  \abstract{We complete the analysis of the effective field theory at the electroweak scale for minimal models of  fundamental partial compositeness. Specifically, we consider fermions in the complex and real representation of the gauge group underlying the composite Higgs dynamics, since the pseudo real representation was investigated earlier. The minimal models feature the cosets SU(4)$\times$SU(4)/SU(4)$_D$ and SU(5)/SO(5) respectively for the complex and real representations. We determine the vacuum alignment, the electroweak precision constraints, additional collider constraints. We finally discuss the main differences among the different models of minimal partial compositeness. 
}

\begin{document}
\maketitle

\pagebreak

\section{Introduction}
The existence of a new, strongly-interacting  dynamics (which we shall denote as technicolor, TC for short) around -- or just above -- the electroweak (EW) scale, has been invoked as a simple way to address  the hierarchy problem of the Standard Model (SM) \cite{Weinberg:1975gm,Susskind:1978ms}. This issue occurs in the SM since there is no natural way to stabilize the Higgs mass against potentially large quantum corrections stemming from unknown physics emerging before the Planck scale. Certain underlying TC theories can be arranged to feature composite dilaton-like Higgs (CDH) \cite{Sannino:2004qp,Dietrich:2005jn} states  or composite Goldstone Higgs (CH) in which the Higgs emerges as  a composite pseudo Nambu-Goldstone Boson (pNGB) \cite{Kaplan:1983fs,Kaplan:1983sm}. The dilatonic-like composite Higgs is expected to emerge near a quantum phase transition\footnote{For controllable perturbative realisations of walking dynamics we refer to \cite{Antipin:2011aa} and references therein. A similar analysis reappeared in \cite{Benini:2019dfy} for a model whose complete phase diagram to the maximum known order in perturbation theory was presented first in \cite{Hansen:2017pwe}. It is worth mentioning that  in \cite{Antipin:2011aa} even the extremely low energy dynamics emerging when chiral symmetry breaks can be exactly investigated.  } \cite{Dietrich:2005jn}, typically required for walking dynamics while the CH  arises from the dynamical breaking of a fundamental fermionic symmetry.  We concentrate on the CH limit although some of our results and scenarios are equally applicable to the CDH limit as well. 

It is well known that in TC the introduction of SM fermion masses and couplings to the Higgs is challenging. Several proposals include extra dimensional constructions while others are based on four-dimensional dynamics. Let us start by considering the following two scenarios for SM fermion mass generation:  
\begin{itemize}
\item[1.] \emph{bilinear coulings}: the composite Higgs is coupled to bilinear operators made out  of the SM fermions;
\item[2.] \emph{partial compositeness}: the elementary SM fermions are linearly coupled (i.e. have a mass mixing) to some composite, non-chiral\footnote{The partners can be given a Dirac mass preserving the SM gauge symmetry.},  heavy partners carrying appropriate quantum numbers. Such partners are assumed to be bound states of the TC dynamics similar to the QCD baryons. Although, as we shall see below, the composite heavy fermions can emerge also from a non QCD-like dynamics. 
\end{itemize}
By construction partial compositeness (PC) requires the introduction of  new QCD coloured states in order to build the composite fermions able to give masses to  quarks.  One observes that a  larger compositeness fraction generically leads to a heavier mass for the SM fermions since the physical eigenstate will be mostly aligned in the direction of the heavy partner. This is a useful guide when trying to generate the observed SM fermion masses including their (intra) family hyerarchies.

In recent years, considerable effort has been made, to classify ultra-violet (UV) theories of PC via gauge interactions featuring only fermionic matter content~\cite{Ferretti:2014qta, Ferretti:2016upr}. 
It remains to be seen whether a purely gauge-fermionic underlying realization of PC trulys exists. One long-standing problem  is the possibility of generating sufficiently large composite fermion anomalous dimensions required to yield the correct top mass and to be larger than the fermion bilinear itself.  This is not possible to achieve within calculable IRFP  theories \cite{Pica:2016rmv} as further confirmed in \cite{BuarqueFranzosi:2019eee}. Even allowing for such a possibility to occur it remains  a daunting task to  generate the observed hierarchies among the SM generations including  the intra-generation splitting. 

It is therefore reasonable and timely to explore composite frameworks in which, while still insisting on the composite nature of the Higgs sector and fermion mass generation, one puts aside (postpone) the naturalness argument.  This  frees us to consider wider classes of composite theories featuring, for example, also TC gauge scalars. It allows for interesting model building featuring unexplored dynamics that can  be investigated via first principle lattice simulations \cite{Pica:2009hc, Chivukula:2010xz, Sinclair:2014cga, Pica:2016zst}. For the scalar-phobic reader we notice that our models can be viewed as an intermediate effective realization of  a more fundamental gauge-fermion dynamics as  explained in \cite{Cacciapaglia:2017cdi}.   TC theories featuring a SM-like Higgs doublet  have been considered long time ago \cite{Simmons:1988fu,Carone:1992rh}, and within the context of CH they  have been re-investigated recently  in~\cite{Antipin:2015jia, Galloway:2016fuo, Agugliaro:2016clv}  to overcome  issues related to the Yukawa sector. 

Within the PC framework an alternative approach has been proposed in~\cite{Sannino:2016sfx}, and further investigated in Ref.s~\cite{Cacciapaglia:2017cdi, Sannino:2017utc}. It was termed \emph{Fundamental Partial Compositeness} (FPC).  The additional ingredient of FPC is the introduction of fundamental techni-scalars, with QCD colour embedded in the corresponding flavour symmetry. The TC-scalars must be chosen in such a way to ensure that Yukawa couplings involving  a SM fermion, a TC fermion, and a TC scalar can be accounted for: this is sufficient to guarantee the generation of fermion masses and Yukawa couplings at low energy. The composite baryons -- i.e. the partners -- are simply made out of one TC-fermion and one TC-scalar, and a UV Lagrangian can straightforwardly be written down, without the need of calling upon the existence of extra coloured TC-fermions. By construction no large anomalous dimensions are needed and the hierarchy among the SM fermions can be simply achieved. Additionally the new fundamental Yukawa structure allows for unexplored flavour dynamics.   Composite theories including (super) coloured TC scalars, attempting to give masses to some of the SM fermions, appeared earlier in the literature \cite{Dobrescu:1995gz,Kagan:1994qg,Altmannshofer:2015esa}  but did not feature a pseudo Nambu Goldstone Boson Higgs.

So far, only the minimal model of FPC (MFPC) has been studied in much depth~\cite{Cacciapaglia:2017cdi}. The latter is based on an $\SU(2)_\text{TC}$ TC gauge group with four (Weyl) TC-fermions in the (pseudo-real) fundamental representation. Without the underlying theory for giving masses to the SM fermions this model was investigated in \cite{Cacciapaglia:2014uja} at the effective Lagrangian level including the  connection to first principle lattice results.  The model features an $\SU(4)/\SP(4)$ coset structure. The aim of the present paper is to extend these analyses to $\SO(N)_\text{TC}$ and $\SU(N)_\text{TC}$ gauge groups with TC-fermions in  real and complex representations, respectively. The TC-scalars are taken to transform according to the fundamental representation of TC. In particular, we  analyse the minimal cases for each class, namely the $\SU(5)/\SO(5)  $ coset for the real case, and $\SU(4)\times \SU(4)/\SU(4) $ coset for the complex one. As these models also feature $\SU(2)_{L/R}$-triplet pNGBs, it is necessary to consider whether they acquire a vacuum expectation value (VEV). The latter, if present, would result in a  violation of the custodial $\SU(2)_L \times \SU(2)_R$ symmetry yielding a contribution to the $T$-parameter. We find that this is indeed the case for $\SU(5)/\SO(5)  $, while all triplet VEVs vanish for $\SU(4)\times \SU(4)/\SU(4) $. As we will show in more detail, this result can be traced back to the different transformation properties under CP-parity of the triplets in the two models under consideration. We also show that the effects of the triplet VEV occur at the order $\mathcal O(p^4)$ in the chiral expansion and therefore there could be an unforeseen cancellation emerging once the coefficients of these operators will be fully determined from the fundamental theory.

The paper is organized as follows: In Section~\ref{sec:fund} we briefly introduce the main features of the fundamental partial compositeness scenario, describing the underlying UV Lagrangian and the associated global and local symmetries. In Section~\ref{sec:real} we move on to discuss the case with TC-fermions in  the real representation, resulting in the $\SU(5)/\SO(5)$ coset. Via the spurion formalism at the effective field theory level, we explore the low energy Yukawa sector. We endow the third generation quarks with masses and further determine their linear couplings to the pNGBs. We then analyse the effective pNGB potential and determine the alignment of the vacuum. Finally, we consider the contribution to sensible EW precision observables such as the $\rho$ parameter and the $Z\bar b b$ coupling, together with further collider constraints, which allow us to restrict the parameter space of the theory.

In Section~\ref{sec:complex} we repeat the above analysis for the case with TC-fermions in a complex representation, yielding the $\SU(4)\times\SU(4)/\SU(4)_D$ coset. In the appendix we classify the various effective operators emerging at different orders in their mass dimension.

\section{Fundamental Lagrangian} \label{sec:fund}

The simplicity of FPC relies in the fact that  enables us to write down a UV theory that, at low-energies, can  reproduce all the experimentally successful predictions of the SM while remaining a valid composite alternative to the SM Higgs sector. In contrast to a purely fermionic realization of PC, no additional QCD colored sectors are required, making the model complete on its own. The Lagrangian structure pertaining to the TC-sector assumes the form:
\begin{align} \label{eq:TClag}
\mathcal L_\text{TC} = \ - \frac 1 4 \mathcal G_{\mu \nu} \mathcal G^{\mu \nu}  +  i { \mF^\dagger} \bar \sigma^\mu D_\mu \mF - \left(\frac 1 2 \mF m_\mF \mF + \text{h.c.} \right) + (D_\mu S) ^\dagger D^\mu S - S^\dagger m_S ^2 S - V(S)\,, 
\end{align}
where $\mathcal G_{\mu \nu}$ refers to the TC gauge fields, $\mathcal F$ is the left-handed Weyl TC-fermion multiplet, $S$ collectively indicates the (complex) TC-scalars, and $V(S)$ is the scalar potential. To keep the notation light we omitted  the TC-indices that are properly contracted to form TC-singlet operators. Up to SM interactions and super-renormalizable operators the symmetry of the Lagrangian~\eqref{eq:TClag} takes the form of a direct product of the fermionic and scalar global symmetries, $G_F \times G_S$.  A diagonal fermion mass matrix breaks the global symmetry of the fermionic kinetic term while a diagonal scalar mass matrix keeps the scalar global symmetry unspoiled. The EW gauging, color interactions and the operators needed to generate the SM fermion masses break the TC global symmetry. 

One can envision the underlying TC fermion matter to transform according to either the real, pseudo-real or complex representation of the TC gauge group. The most minimal realization of FPC in terms of the needed fields is, as explained in~\cite{Cacciapaglia:2014uja,Sannino:2016sfx,Cacciapaglia:2017cdi},  the case of $\SU(2)_\text{TC}$ with fields in the fundamental representation, which for this gauge group is a pseudo-real representation. 

For both the pseudo-real and real TC-color representations\footnote{Here we  always choose TC matter to be in a  defining representation of the gauge group  which will therefore be either  Sp$(2N)_{TC}$ or SO$(N)_{TC}$ with SU$(2)_{TC}$=Sp$(2)_{TC}$ being the first of the symplectic groups.} we can arrange the field $S$ in a multiplet $\Phi$ as follows
\begin{align} \label{eq:grouping}
\Phi =\left( \begin{array}{cc} S \\ \overline S \end{array} \right)\,, \qquad  
\overline S =\begin{cases}
 \epsilon_\text{TC} S^* & \SP(2N_s) \\
 \quad S^* & \SO (2N_s) 
\end{cases}
\end{align}
where on the right we indicated the corresponding global symmetries over the scalars, which are $\SP(2 N_S)$ or $\SO(2N_S)$ depending on whether the TC-color gauge representations are pseudo-real or real. The quadratic scalar Lagrangian reads:
\begin{align}
\frac 1 2 (D_\mu \Phi )^T \left( \begin{array}{cc} &  \ \pm1 \\   1 & \end{array} \right)  (D_\mu \Phi ) - \frac 1 2 \Phi^T \left( \begin{array}{cc} & \pm {m_S^2}^T \\ m_S^2 & \end{array} \right)  \Phi\,,
\end{align}
where the plus(minus) sign corresponds to $\SO(2N_S)$($\SP(2N_S)$) and we introduce the off-diagonal matrix $\displaystyle{\omega = \left( \begin{array}{cc} &  \ \pm1 \\   1 & \end{array} \right)}$ since we will be using it later. 
Note that we  need  the  global symmetry over the scalars to be at least  $\SU(3)$ to account for QCD color, since the TC-fermions are taken to be color singlets. 
In the complex TC-color representation the maximum scalar symmetry is $\U(N_S)$. 

Since the symplectic case has been studied in much detail in \cite{Cacciapaglia:2017cdi}, in the following we concentrate  on the real and complex representations.

\section{Real case} \label{sec:real}
In this Section we consider the case of an $\SO(N)_\text{TC}$ gauge group. As has been observed, the minimal coset that contains a pNGB Higgs with a custodial symmetry, and with TC-fermions in the fundamental, is $\SU(5)/\SO(5)$. The latter coset has been explored in different contexts of possible UV completions  of composite Higgs theories, e.g., in Ref.s~\cite{Agugliaro:2016clv, Ferretti:2016upr}. We  investigate here the high energy extension of the $\SU(5)/\SO(5)$ composite model assuming the FPC scenario. 
\begin{table}[t]
\begin{tabular}{ l c  c  c  c } 
\hline 
 states & $\SO(\ntc)$ & $\SU(N_F)$ & $\SO(2 N_S)$ & number of states \\ \hline
 $\mF $ & $\ntc$ &  \tiny{ \yng(1)} & 1 & $\ntc \times N_F$ \\ 
  $\Phi$ & $\ntc$ &  1 &   \tiny{ \yng(1)} &  $\ntc \times 2 N_S $ \\ \hline
  $\Phi \Phi$ & 1 & 1 & $ \Yvcentermath1 1\oplus \tiny{ \yng(2)}\oplus \tiny{ \yng(1,1)}  $ & $ 1 + 2N_S(2N_S+1) + N_S(2N_S-1)-1 $ \\ 
  $\mF \Phi$ & 1 & \tiny{ \yng(1)}  &   \tiny{ \yng(1)}  & $2 N_S \times N_F$ \\
  $\mF \mF$ & 1 & $1\oplus \tiny{\yng(2)}$ & 1 &$ \frac 1 2 N_F( N_F+1) + 1 $  \\ \hline
\end{tabular}
\caption{Fundamental and composite matter fields for  $\gtc=\SO(\ntc)$.  $N_F$ ($N_S$) is the number of TC-fermions (scalars).}
\label{tab:tab1}
\end{table}
\subsection{Details of the model}
The underlying fermionic theory features 5 Weyl fermions carrying the following quantum numbers under the SM group $G_\text{SM} = \SU(3)_c \otimes \SU(2)_L \otimes \U(1)_Y$
\be \label{eq:fundferm}
\mF_\pm \equiv \left(1, 2, \pm \frac 1 2 \right), \ \mF_0 \equiv \left(1, 1, 0\right)\,.
\ee

The electroweak symmetry is embedded inside a larger custodial $\SU(2)_L \times \SU(2)_R$ group, identified with the following six $\SU(5)$ generators:
\be\label{su5:gen}
T_L ^i = \frac{1}{2}
\left(
\begin{array}{c | c} 
\mbox{$ \mathbb{1} _2 \otimes \sigma^i $}  &   \\
\hline
  & 0 \end{array} \right)\,,
\quad 
T_R ^i = \frac{1}{2}
\left(
\begin{array}{c | c}
\mbox{$ \sigma^i \otimes \mathbb{1}_2 $ } & \\
\hline
& 0 \end{array} \right)\,.
\ee
With such an embedding, the TC-fermions given in eq.~\eqref{eq:fundferm} can be grouped into a single field $\mathfrak{F}$ transforming according to the fundamental of $\SU(5)$
\be
\mathfrak F = \left( \begin{array}{c}
\mF _{+} \\ \mF _- \\ \mF_0  \end{array} \right)\,.
\ee
The fundamental UV Lagrangian for the free $\mathfrak{F}$ field is 
\begin{align} \label{eq:luv}
\mathcal{L}_{\rm UV} &= i \mathfrak{F}^\dagger \overline \sigma^\mu D_\mu \mathfrak{F} - \mathfrak{F}\mathcal{M} \mathfrak{F}+ \mbox{h.c} \nonumber  \\
&= i \mathfrak{F}^\dagger \overline \sigma^\mu D_\mu \mathfrak{F}  - \mathfrak F^T \left( \begin{array}{c c c} & i \sigma^2 \mu_d & \\ - i \sigma^2 \mu_d & & \\ & & \mu_s \end{array} \right) \mathfrak F  + \mathrm{h.c.} \,, 
\end{align}
\be
D_\mu = \partial_\mu - i g W^i_\mu T_L^i - i g' B_\mu T_R^3 - i g_{\rm TC} G_\mu^a \lambda^a_{\rm TC}\,.
\ee
The choice of the condensate $\big \langle \mathfrak F \mathfrak F \big \rangle$ is fixed (modulo an overall phase) by the embedding of the SM symmetry and by the requirement that the stability group be $\SO(5)$. Our choice will be
 \be \label{vacuum}
\Sigma_0=\left(
\begin{array}{cc|c}
 & i \sigma^2 &  \\
 -i \sigma^2 &  & \\ \hline
  & & 1 
\end{array} \right)\,.
\ee
The low-energy dynamics is described in terms of a linearly transforming matrix $\Sigma$, defined as
\be\label{goldstone}
\Sigma(x) = \Omega(\theta) \, \exp \left[4 i \Pi(x) /f \right] \Sigma_0 \, \Omega^\intercal (\theta)\,, 
\ee
where $\Pi (x) = \Pi^{\hat a} (x) X^{\hat a}$ is the pion matrix, $X^{\hat a}$ are the $\SU(5)$ generators broken by the vacuum $\Sigma_0$, and $\Tr [X^{\hat a} X^{\hat b}] = \frac{1}{2} \delta^{\hat{a} \hat{b}}$, while $f$ is the physical scale generated by the strong TC-dynamics. The rotation matrix $\Omega(\theta)$ describes the misalignment of the vacuum from the EW preserving direction. Note that the mass matrix in eq.\eqref{eq:luv} is proportional to $\Sigma_0$ only for $\mu_d = \mu_s$, and that the custodial $\SO(4) \subset \SO(5)$ symmetry is always preserved by the mass term.  

For the explicit form of the pion matrix we refer the reader to  Ref.~\cite{Agugliaro:2018vsu}.  The Goldstones include a custodial bi-triplet $(3,3)$, which under the diagonal symmetry $\SU(2)_D$ they decompose as
\be
{(3,3)} \rightarrow  {\bf 5} \oplus {\bf 3} \oplus {\bf 1}\,,
\ee
and the corresponding fiveplet, triplet and singlet states will be denoted as follows:
\be
{\bf 5} = ( \eta_5^{++},\ \eta_5^+,\ \eta_5^0,\ \eta_5^-,\ \eta_5^{--})\,, \quad {\bf 3} = (\eta_3^+,\ \eta_3^0,\ \eta_3^-)\,, \quad {\bf 1} = \eta_1^0\,.
\ee
In the FPC scenario we are considering here, in addition to the light pNGB states,  we expect massive states which, differently from the purely TC-fermionic case, will include TC bound states containing either only TC-scalars or a TC-scalar and a TC-fermion. For the reader convenience we summarize in Table~\ref{tab:tab1} the elementary states as well as the first scalar and fermionic resonances together with their transformation properties under the global symmetries.
\subsection{Yukawa interactions}
Following  \cite{Cacciapaglia:2017cdi}, the fundamental Yukawa interactions can be put in a formally invariant way by introducing a spurion $\psi^i\,_a$, where the index $i$ refers to the symmetry of the scalar sector -- of which QCD color is a subgroup -- while $a$ is the index referring to the fundamental of $\SU(5)$. The transformation properties of $\psi^i \,_a$ under $\SO(2N_S)\times \SU(5)$ are schematically summarized by
\begin{align}
&\psi^i \, _a  \in   \ytableausetup
{aligntableaux=top, boxsize=.7em}  {{\begin{ytableau}[]  \\ \end{ytableau}}}_S \otimes \overline{{\begin{ytableau}[]  \\ \end{ytableau}}}_\mF\,. 
\end{align}
Using this spurion notation, the partial fundamental Yukawa Lagrangian  preserving the flavour symmetries can be compactly  written as
\be \label{eq:fundyuk}
\mathcal L_{\text{Yuk}} = - \ \psi^i \, _ a \Phi^j  \omega_{ij} \mF^a +\text{ h.c.}\,
\ee
where the TC indices have been omitted, being implicitly contracted to yield a TC-singlet. Notice that, while two upper $i$ indices must be contracted via $\omega_{ij}$  upper and lower indices are contracted simply  with a $\delta^i _{j}$, since $\Phi^T \omega \Phi = \Phi^\dagger \Phi$.


The partial fundamental Yukawa Lagrangian that allows to give mass   to the top and to the bottom quarks is
\begin{align} \label{eq:yukson}
\mathcal L _{\text{quark}}\ = \ & 
y_Q \mF_+ ^\alpha S_t Q_\alpha +\tilde y_Q  \mF_- ^\alpha S_{b} Q_\alpha + y_{t} \mF_0 S_t ^* t_R ^c + y_{b} \mF_0 S_{b} ^* b_R ^c,  \\\nonumber \\
&  S_t \equiv \left(\bar 3, 1, - \frac 2 3\right), \ S_{b} \equiv \left(\bar 3, 1,  \frac 1 3\right)\,,
\end{align}
where we split the original scalar multiplet into two different  color-triplet techniscalar fields, $S_t$ and $S_b$, which implies that $N_S=  6$, up to lepton scalars and the remaining quark generations. Following the argument made above,  the largest possible scalar symmetry, so far, is $\SO(12)$, which is realized when $S_t$ and $S_b$ are degenerate and the hypercharge interactions are switched off; conversely, in the presence of a large splitting one has simply $\SO(6)\times \SO(6)$. In the following we will assume all the scalars to have common masses that are light with respect to the dynamical scale of the TC theory. Therefore the common composite scale will be of the order of the dynamical TC scale. This also means that the Yukawa couplings of the fundamental theory must contain the hierarchy information to reproduce the SM fermion mass hierarchy. 

By comparing eq.\eqref{eq:yukson} and eq.\eqref{eq:fundyuk}, the explicit expression of the matrix field $\psi^i \, _a$ is found to be
\begin{align}
\psi^i \,_a = \ \left(
\begin{array}{ccccc}
 0 & 0 & 0 & 0 & y_t  t_R^c \\
 0 & 0 & 0 & 0 & y_b b_R^c \\
 -y_Q b_L & y_Q t_L & 0 & 0 & 0 \\
 0 & 0 & -\tilde y_Q   b_L & \tilde y_Q t_L & 0 \\
\end{array}\right)\,.
\end{align}
At lowest order, the only invariant operator is bilinear in the SM fermions, and describes the masses of the quarks and their interactions with the pNGBs. This is given by
\begin{align} \label{eq:yuksu5}
\mathcal O _{\text{Yuk}} = \ & i \frac {f} {2\sqrt{2}} \left(\psi^{i_1}\, _{a_1} \psi^{i_2}\,_ {a_2}\right) \Sigma^{a_1 a_2} \omega_{i_1 i_2}\,.
\end{align}
Expanding the above operator to linear order in the pNGBs, we get the following Yukawa Lagrangian for the neutral fields
\begin{align} \label{quarkmass}
\mathcal L_{\text{Yuk}} ^0  = \ & C_\text{Yuk} \mathcal O_\text{Yuk} = \nonumber \\
  =\ & - C_{\text{Yuk}}    y_Q y_t  \left[ v \ct+ h \ctt - i \st\ct\left(\frac{3}{\sqrt{10}}\eta - \sqrt{\frac{3}{2}} \eta_1 ^0 \right) + \st \, \eta_3^0 \right] t_L t_R^c  \nonumber \\
 \ & - C_{\text{Yuk}}   \tilde y_Q y_b  \left[ v \ct+ h \ctt - i \st\ct\left(\frac{3}{\sqrt{10}}\eta - \sqrt{\frac{3}{2}} \eta_1 ^0 \right) - \st \, \eta_3^0 \right] b_L b_R ^c\,,
\end{align}
and for the charged fields
\begin{align}
\mathcal L_{\text{Yuk}} ^{\pm}= \ & C_\text{Yuk}\left(- \sqrt{2} y_Q y_t b_L t_R ^c  \eta_3 ^+  - \sqrt{2} \tilde y_Q  y_b  t_L b_R ^c  \eta_3 ^-  + \text{h.c.} \right)\,. \
\end{align}
From \eqref{eq:yuksu5} one can easily read off the masses of the top and bottom quarks
\begin{align} \label{eq:quarkmass}
m_t = \lvert C_\text{Yuk} \, y_Q y_t| v \ct, \quad m_b = \lvert C_\text{Yuk} \, \tilde y_Q y_b| v \ct \,.
\end{align}

So far the SM mass hierarchy was encoded in the fundamental Yukawas, however it could also be induced by the scalars possessing different masses. In this case we expect the TC scalars to have masses heavier than $\Lambda$. In this case we can integrate the scalars out. At the effective Lagrangian level this amounts to replacing the $\omega_{ij}$ tensor with~\cite{Cacciapaglia:2017cdi}
\be
\omega_{ij} \to \left( \frac{\Lambda ^2}{M_S ^2} \right)_{ij} \,.
\ee
%
\vspace{.1cm}

\subsection{Partners and exotic states}
As already discussed, the  top and bottom quark partners are composite states made out of one TC-scalar and one TC-fermion. Of course, in addition to the SM quark partners, also exotic vector-like quarks (VLQs) ${\cal B} \sim {\cal F} \Phi $ arise, whose decays lead to interesting collider signatures.  
According to Table~\ref{tab:tab1}, these states carry one index in the fundamental of $\SO(12)$ and one in the fundamental of $\SU(5)$. Therefore,  fermionic partners will be found in the $\mathbf 5$ of $\SU(5)$ times a $\mathbf 3$ or a $\bar  {\mathbf 3}$ of color, and their  hypercharge will be equal to the sum of those of the scalar  and fermionic fields, $S$ and $\mathcal F$,  they are constituted of. Calling $Y_S$ the hypercharge associated with the scalar field $S$, the decomposition of the partners under the EW group thus reads
\begin{align}
  \mathbf{2}_{1/2 +Y_S} \oplus \mathbf 2_{-1/2+Y_S} \oplus \mathbf 1_{Y_S}\,.
\end{align}
For ease of reading  we summarize the SM quantum numbers of the partners in Table~\ref{tab:partner_dec}, where each fermion is left-handed, and the corresponding VLQ fields are formed by grouping together the tilded and non-tilded fields to form Dirac fermions.
\renewcommand{\arraystretch}{1.7}
\begin{table}
\centering
\begin{tabular}{c | c c c}\hline 
${\cal{B}}^i \, _a$ & $\mF_+$ & $\mF_-$ & $\mF_0$ \\  \hline
  $S_t^*$ & $  {\cal{B}}_X\,(3, 2)_{7/6}$ & $  {\cal{B}}_Q \, (3,2)_{1/6}   $ & $  {\cal{B}}_t \,(3, 1)_{2/3}$ \\ 
  $S_b^*$ & $ {\cal{B}}_Q ' \,(3, 2)_{1/6}$ & $  {\cal{B}}_Y \, (3,2)_{-5/6}$ & $ {\cal{B}}_b\, (3, 1)_{-1/3}$ \\ 
 $S_t$ & $\widetilde {\cal{B}}_Q\,  (\overline 3, 2) _{-1/6}   $ & $ \widetilde  {\cal{B}}_X \, (\overline 3, 2)_{-7/6}$ & $\widetilde  {\cal{B}}_t\, (\overline 3, 1)_{-2/3}$ \\ 
  $S_b$ & $ \widetilde {\cal{B}}_Y\, (\overline 3, 2)_{5/6}$ & $\widetilde  {\cal{B}}_Q ' \,(\overline 3, 2)_{-1/6}$ & $\widetilde  {\cal{B}}_b \,(\overline 3, 1)_{1/3}$ \\ 
  \hline
\end{tabular}
\caption{Representations of the composite partners under the SM gauge group. The rows and columns correspond to the fundamental TC-scalar and TC-fermion constituents, respectively.}
\label{tab:partner_dec}
\end{table}

 
The resulting effective Lagrangian mixing SM fermions and TC composite fermions, including the Dirac masses for the latter, is
\begin{align} \label{eq:partners}
& -\, \mathcal L_\text{mix} = \   f y_i^\text{eff}  \, \omega_{ij} {\cal{B}}^i\,_a \Sigma ^{ab} \psi^j\,_b  +  M_i  \, \omega_{ij}{\cal{B}}^i\,_a \Sigma^{ab} {\cal B} ^j\,_b \, ,
\end{align}
where ${\cal{B}}^i\,_a$ is the matrix of partners of Table~\ref{tab:partner_dec}.

By expanding the above expression we determine the top quark mass, that at leading order in the chiral expansion reads:
\begin{align}
m_t = \  \sqrt 2 M_T \st \ct \frac{y_Q ^\text{eff} f}{\sqrt{M_T^2 + (y_Q^\text{eff})^2 f^2}}\frac{y_t ^\text{eff} f}{\sqrt{M_T^2 + (y_t^\text{eff})^2 f^2}} + \dots 
\end{align}
Note that the above expression matches with eq.~\eqref{eq:quarkmass}, after the correct identifications have been made, as explained in Ref.~\cite{Cacciapaglia:2017cdi}. Here the reader will find also the relations, in certain limits, between the effective Yukawas and the fundamental Yukawas.  
We leave to future studies the decay and phenomenology of exotic partners.

\subsection{Potential and vacuum misalignment} 
We now discuss the effective potential for the composite fields, paying attention to those  that are relevant for the breaking of the  electroweak symmetry. One can naively imagine these operators to  arise 
 via  loops of  top partners.  
The potential contains three different  contributions, corresponding to the various sources of explicit global TC-fermion symmetry breaking. The pGNB potential term stemming from one loop of  SM gauge bosons reads
\begin{align}
V_{\text{gauge}} &= \ C_g  g^2 f^4 \Tr\left[T_L ^i \Sigma \, {T_L^i}^* \Sigma^*\right]+ C_g  {g'}^2 f^4 \Tr \left[T_R ^3 \Sigma \, {T_R^3}^* \Sigma^*\right] = \nonumber \\
&=  C_g f^4 \frac{3g^2 + g'^2}{2}\left( -(\ctt +1)+ 2 \stt \frac h f + \dots    \right) \ .
\end{align}
A bare mass term for the TC fermions leads to
\begin{align}
V_{\text{mass}} &= \ - C_m f^3 \Tr[M_{{\cal F}} \Sigma + \Sigma^\dagger M_{{\cal F} }] = \nonumber \\
&=  2 C_m f^3 \left( -3 \mu_d - (\mu_d + \mu_s) \ctt + 2 (\mu_d + \mu_s) \stt \frac h f + \dots \right)\, \, ,
\end{align}
where $M_{{\cal F}}$ is the following matrix
\begin{align}
M_{{\cal F}} = \left( \begin{array}{ccc} 0 &\mu_d (i\sigma^2) & 0 \\ - \mu_d (i\sigma^2) & 0 & 0 \\ 0 & 0 & \mu_s \end{array} \right) \,.
\end{align}
Note that, as long as $C_g, C_m >0$, both the gauge and mass term tend to align the vacuum in the unbroken direction, corresponding to $\theta=0$.

One further source of explicit breaking is provided by the fundamental Yukawa couplings in eq.~\eqref{eq:fundyuk}. In order to write down the potential generated by the Yukawa couplings, we  extract the associated spurion matrix $y$  defined by
\begin{align} \label{eq:fermdec}
& \spur{i}{a}  = (\Psi \, y )^i \, _a,  \ (y_\alpha)^i \,_a \in \tiny{ \yng(1)}_S  \otimes \overline {\tiny{ \yng(1)}}_{\cal{F}} \,,\end{align}
where $\Psi$ is a generic SM fermion. Therefore, if a  fermion $f$ transforms according to the representation $R_\text{SM}$ of the SM gauge group, the associated Yukawa spurion $y_f$  will have the following transformation properties
\begin{align}
(y_f) ^i \, _a  \in \tiny{ \yng(1)}_S  \otimes \overline {\tiny{ \yng(1)}}_{\cal{F}}   \otimes{ \overline{ R} } _\text{SM}  \,.
\end{align}

The potential is made by the following three operators, $O_V ^{1,2,3}$ \cite{Cacciapaglia:2017cdi} (cfr. eq.s~(38--40))
	\begin{align}
	\mathcal{O}^1_{V_f} = \dfrac{f^2 \Lambda^2 }{16 \pi^2} (y_f^\ast y_f)^{a_1} \phantom{}_{a_2} \phantom{}^{i_1}\phantom{}^ {i_2} (y_{f'}^\ast y_{f'})^{a_3} \phantom{}_{a_4} \phantom{}^{i_3}\phantom{}^ {i_4}  \Sigma^{\dagger}_{a_1 a_3} \Sigma^{a_2 a_4} \omega_{i_1}\phantom{}_{ i_2} \omega_{i_3}\phantom{}_{ i_4}\ , \label{eq:OVf1}\\
	\mathcal{O}^2_{V_f} =  \dfrac{f^2 \Lambda^2 }{16 \pi^2} (y_f^\ast y_f)^{a_1} \phantom{}_{a_2} \phantom{}^{i_1}\phantom{}^ {i_2} (y_{f'}^\ast y_{f'})^{a_3} \phantom{}_{a_4} \phantom{}^{i_3}\phantom{}^ {i_4}  \Sigma^{\dagger}_{a_1 a_3} \Sigma^{a_2 a_4} \omega_{i_1 i_3} \omega^{i_2 i_4}\ , \label{eq:OVf2}\\
	\mathcal{O}^3_{V_f} =  \dfrac{f^2 \Lambda^2 }{16 \pi^2} (y_f^\ast y_f)^{a_1} \phantom{}_{a_2} \phantom{}^{i_1}\phantom{}^ {i_2} (y_{f'}^\ast y_{f'})^{a_3} \phantom{}_{a_4} \phantom{}^{i_3}\phantom{}^ {i_4}  \Sigma^{\dagger}_{a_1 a_3} \Sigma^{a_2 a_4} \omega_{i_3}\phantom{}_{ i_2} \omega_{i_1}\phantom{} _{i_4}\ . \label{eq:OVf3}
	\end{align}    
Adding the gauge, TC-fermion masses, and Yukawa  the potential for $\theta$  is 
\begin{align}
V = \ &f^4\left(-A \cos 2\theta + B \cos 4 \theta\right), 
\end{align}
where the coefficients $A$ and $B$ are given by
\begin{align}
A = \ & C_g  \frac{3g^2 + g'^2}{2} + 2 \, \frac{C_m} {f}  (\mu_d + \mu_s) + \frac{3 \Lambda^2}{16 \pi^2 f^2} \left[ \frac {3} 2 C_{V_f} ^1 \left(|y_Q|^2+|\tilde y_Q|^2\right) \right.  \nonumber \\
 & \left. + \, C_{V_f}^3 \left( 2 |y_Q|^2|\tilde y_Q|^2 + \frac 1 2 (|y_Q|^4+|\tilde y_Q|^4) \right) \right] \,, \\
 B = \ & \frac{3 \Lambda^2}{16 \pi^2 f^2} \left[ - \frac {1} 2 C_{V_f}  ^2  \left( |y_Q y_t|^2 +|\tilde y_Q y_b|^2 \right) + \frac {3} 8 C_{V_f}^1 \Big[ 2(|y_b|^2+|y_t|^2)  - (|y_Q|^2+|\tilde y_Q|^2)\Big]^2 \right. \nonumber \\
   & \left. + \, \frac {1} 8  C_{V_f}^3 \left( 4(|y_b|^4+|y_t|^4)+ (|y_Q|^4+|\tilde y_Q|^4)\right) \right] \,.
\end{align}
We observe that the structure of the vacuum-potential as a function of  $\theta$ reproduces  that found in the case of  {fermion partial compositeness} studied in~\cite{Agugliaro:2018vsu}. Also note that for the electroweak symmetry to be spontaneously broken we need $A, B >0$.  

\subsection{Triplet tadpole}
A common issue in non-minimal models is the possibility for the weak gauge triplets  pNGBs to acquire a vacuum expectation value. For the $\SU(5)/\SO(5)$ coset, as discussed in detail in~\cite{Agugliaro:2018vsu}, the CP-even neutral custodial triplet field, $\eta_3 ^0$, can acquire a VEV induced by the following tadpole term
\begin{align} \label{eq:tadpole}
V_y \supset \ \frac{4 f \Lambda^2}{16\pi^2 } \eta_3^0 \ct \st^2 & \left[ -  C_{V_f}^1 (|y_Q|^2-|\tilde y_Q|^2) (|y_Q|^2+ |\tilde y_Q|^2-2(|y_b|^2+|y_t|^2)) \right. \nonumber \\
& \left. + C_{V_f}^2 \left( |y_Q y_t|^2 - |\tilde y_Q y_b|^2\right) + C_{V_f}^3 \left(|\tilde y_Q|^4-|y_Q|^4\right)\right]\,.
\end{align}
Barring cancellations between the different operators, the above expression vanishes only in the custodial limit $ y_t \to y_b, \, y_Q \to \tilde y_Q$. In this limit  the  $\SU(2)_R$ symmetry is preserved by the fundamental Yukawa sector, with the top and bottom quarks being degenerate. Hence, requiring physical masses inevitably induces a tadpole for the $\eta_3^0$. Note that the degree of tuning that we must demand is dictated by the ratio between the bottom and top masses given in eq.~\eqref{eq:quarkmass}: 
\be \label{eq:finetune}
\frac{m_b}{m_t} = \Bigg \lvert \frac{\tilde y_Q y_b} {y_Q y_t}\Bigg \rvert \approx 2 \, \%\,.
\ee

Let us now compute the contribution of the triplet tadpole to the $\rho$ parameter. The field $\eta_3^0$ has the following tadpole and mass term
\begin{align}
V_{\eta_3 ^0} = f^3 \ct \st^2 T_\eta \eta_3 ^0 + \frac 1 2  \left(m_{\eta_3^0}^2+ \st^2\delta m_{\eta_3^0}^2\right) (\eta_3^0)^2 \,.
\end{align}
Solving the equation of motion for $\eta_3^0$, and retaining only the leading term in $\st^2$, we find its VEV
\be
\big \langle \eta_3^0 \big \rangle = - \frac{ f^3 T_\eta \st^2}{2 m_{\eta_3^0}^2}\,, 
\ee
resulting in  the following contribution to the $\rho$ $(\equiv m_W^2\big/ m_Z^2 \cos^2 \theta_W)$ parameter, up to   $\mathcal O(\st^4)$ corrections
\be
\delta\rho \Big |_{\eta_3^0} =  - \frac{ 2 f^4 T_\eta ^2  \st^2}{m_{\eta_3 ^0} ^4}  \,.
\ee
This contribution should be added to those coming from the four-fermion operators of eq.s~(\ref{4f:first}--\ref{4f:last}), and the full contribution is given in eq.~\eqref{eq:deltarho}. 
For completeness, we also report the mass of the neutral triplet, obtained by expanding $V_g$, $V_m$ and $O_{V_f}^{1-3}$ to the second order in the pNGB field. The result reads 
\begin{align}
m_{\eta_3^0} ^2 \stackrel{}{\approx} 4 C_g f^2  \left( 2g +g'^2 \right)+ 8 C_m f \mu_d-4 f^2  (C_{V_f}^1+C_{V_f}^3) (|y_Q|^4+|\tilde y_Q|^4) + \mathcal O(\st^2)\,.
\end{align}
Note that the $\eta_3^0$ state becomes tachyonic when the effect of the SM fermion loops becomes larger than that due to  EW gauge loops. In general, one can exclude wide regions of parameter space by using the requirement that no tachyons should be present in the spectrum~\cite{Cai:2018tet, Agugliaro:2018vsu}.
\subsection{Corrections to $Z \bar b b$}
The operator that gives corrections to the fermion couplings to gauge bosons has the structure
\begin{equation} \label{eq:OPif}
	\mathcal{O}_{\Pi f} = \frac{i f}{4 \Lambda} \ (\spur{i_1}{a_1} ^\dagger\bar{\sigma}^\mu \spur{i_2}{a_2} )\  \Sigma _{a_1 a_3}^\dag  \overleftrightarrow{D}_\mu \Sigma ^{a_3 a_2}\  \omega _{i_1 i_2}\,.
	\end{equation}	
where $\bar \sigma^\mu = (1, \vec{\sigma})$. By expanding the above operator to the zeroth order in the pNGBs, we find the following corrections to the $Z$ and $W$ fermion couplings
\begin{align}
O_{\Pi f} \ \supset  \ \frac{g}{2\cos \theta _W}  \frac f {\Lambda} \st^2  & Z_\mu  \left( | y_Q |^2 t_L ^\dagger \bar \sigma^\mu t_L - |\tilde y_Q|^2 b_L ^\dagger \bar \sigma^\mu b_L \right) \nonumber \\
& +\frac{g}{2\sqrt{2}}  \frac f {\Lambda} \st^2 W_\mu ^+ \left( |y_Q|^2+|\tilde y_Q|^2 \right)  t_L ^\dagger \bar \sigma^\mu b_L + \text{h.c.}\,
\end{align}
Two observations are in order: i) No corrections to right-handed couplings are generated, since these  couple only to the gauge singlet fermion $\mF_0$. 
This is different from  the $\SU(4)/\SP(4)$ case where both left- and right-handed couplings are generated, since the TC fermions feature no SM singlet. 
ii)  Differently from the pseudo-real case  the  left-handed top and bottom quarks couplings are weighted by two different coefficients, i.e. $|y_Q|^2$ and $|\tilde y_Q|^2$, respectively.

From eq.~\eqref{eq:OPif}, using the best fit value for $\delta g_{Lb}$ determined in~\cite{Gori:2015nqa}, we obtain the bound
\begin{align} \label{eq:zbb}
C_{\Pi f} \lvert \tilde y_Q |^2 \st ^2 <  0.1, \quad @ \, 95\% \ \text{CL.}
\end{align} 
By comparing with eq.~\eqref{eq:quarkmass}, the above constraint translates into 
\begin{align}
\lvert y_b \rvert \frac{ \lvert C_\text{Yuk} \rvert}{\sqrt{\lvert C_{\Pi f} \rvert}} \gtrsim \frac{m_b}{f \ct \sqrt{0.1}} \approx 0.02 \frac{10\TeV}{\Lambda}\,.
\end{align}
The above bound suggests that, unless Naive-Dimensional Analysis (NDA)  -- which would predict $|C_\text{Yuk}| \simeq |C_{\Pi f}|$ -- is apparently violated,  the right-handed bottom mixing $|y_b|$ should be larger than  $2\%$ for condensation scales $\Lambda$ around $10 \TeV$. In turn, this implies that, within NDA, the right-handed bottom coupling is able to accommodate the hierarchy of eq.~\eqref{eq:finetune}. 

\subsection{NLO corrections to the kinetic term}

The  NLO  operators giving corrections to the kinetic term of the non-linear field $\Sigma$ are listed in Appendix~\ref{app:NLOSU5}. Only some of them  contribute to the $\rho$ parameter. Including also the contribution from the triplet tadpole, the correction to the $\rho$ parameter is summarized as
\be \label{eq:deltarho}
\delta\rho = \delta\rho \Big |_{\eta_3^0}  + \frac {3\Lambda^2 \st^2} {4\pi^2 f^2}  \,  C_{y\Pi D} \left( |y_Q|^2 - |\tilde y_Q|^2\right)^2   + \frac{\st ^2  g'^2}{4\pi^2} \, C_{\Pi D} \, \,, 
\ee
where  $C_{y\Pi D}$ and $C_{\Pi D}$ are defined in eq.~\eqref{eq:cypd}, and the common dependence on $\st^2$ has been factored out for each term. We note that eq.~\eqref{eq:deltarho} is a rough estimate of the correction to the $\rho$ parameter. Additional contributions come from  gauge-boson vacuum polarization diagrams, with the heavy vector-like partners running in the loop, and are more important for large mixings, as it is the case for the top-quark~\cite{Xiao:2014kba}.

 \subsection{Effective interactions for the top sector}
Expanding the four-fermion operators in eq.s~(\ref{eq:fourferm1}-\ref{eq:fourferm8}) we can extract  contact interactions among the SM fields. The results are given in eq.s~(\ref{4f:1}-\ref{4f:6}). We are interested specifically in the associated Wilson coefficients expressed in terms of the parameters of the fundamental UV theory, up to form factors generated by the strong TC dynamics.
An interesting observable involves four top quarks in the final state
\begin{align}
\mathcal L _\text{EFT} \supset \  \frac{\ctt^2 C_{4f}^3 +  C_{4f}^4 +  C_{4f}^5}{4\Lambda^2} \big\lvert y_t \big \rvert^4  (\overline t_R \gamma^\mu t_R)(\overline t_R \gamma_\mu t_R)\,,
\end{align}
which has been directly probed experimentally at LHC. The ATLAS bound~\cite{Aaboud:2018xuw} at 95$\%$ C.L. leads to
\be
\frac{|y_t|^4}{4\Lambda^2} \Big\lvert \ctt^2 C_{4f}^3 +  C_{4f}^4 +  C_{4f}^5 \Big\rvert <  2.9 \TeV^{-2}  \ \Rightarrow \  \Big\lvert \ctt^2 C_{4f}^3 +  C_{4f}^4 +  C_{4f}^5 \Big \rvert ^{1/4} |y_t| < 5.8 \left( \frac{\Lambda}{10 \TeV} \right)^{1/2}
\ee
Furthermore, the dipole operators of eq.s~(\ref{eq:dip1}-\ref{eq:dip2}) generate new interactions between gauge fields and SM fermions.
To compute these operators, we recall that the EW and QCD color generators are embedded in the $\SU(5)$ and $\SO(12)$ flavour symmetries,\footnote{Note that, in the case of non-degenerate TC scalar fields, when the scalar symmetry is reduced to $\SO(6)\times \SO(6)$, the embedding of the QCD gauge symmetry remains unchanged.} respectively, as follows
\renewcommand{\arraystretch}{0.9}
\begin{align}
A_\mu ^I T_{F} ^I =& \ \frac 1 2  
\left( \begin{array}{c cc } g W_\mu ^i \sigma^i + g' B_\mu  & 0  & \\
0 & g W_\mu ^i \sigma^i -g' B_\mu & \\ 
& & \ 0 \end{array} \right) \,,    \\
G_\mu ^A T_\mathcal{S} ^A =& \  \frac  {g_s} 2 G_\mu ^a   
\left( \begin{matrix} 
- \lambda_a ^ T & & & \\
& -\lambda_a ^T & &  \\
& & \ \lambda_a  & \\
& & & \ \lambda_a  \end{matrix} \right) +  \frac {g'} 3  B_\mu 
 \left( \begin{array}{cccc}  -2   & & & \\ &    1 & & \\ & & 2   & \\ & & & -    1 \end{array} \right)\,.
\end{align}
Here $\lambda_a$ are the SU(3) Gell-Mann matrices  and $g_s$ is the strong coupling.
\vskip .2cm

The dipole operators~(\ref{eq:dip1}-\ref{eq:dip2}) generate the following couplings:
\begin{align}
\mathcal O _{fW} &= \ \frac{- m_t}{C_\text{Yuk} \Lambda^2 \, v} \cdot \frac{\ctt}{\ct} \left( g \mathcal O_{uW}^{33 *} + g' \mathcal O_{uB}^{33 *}  \right) + \ \dots	  \\
\mathcal O _{fG} &= \ \frac{- 2 \,m_t}{C_\text{Yuk} \Lambda^2 \, v} \cdot \frac{\ctt}{\ct}  \, g_s \mathcal O_{uG}^{33 *} + \ \dots
\end{align}
where the operators $\mathcal O_{uV} ^{33}$ come from the SM EFT~\cite{PhysRevD.91.074017}, and the dots contain higher order interactions generated by the non-linearities,   involving multiple pNGBs.

The constraints extracted from the \texttt{TopFitter} collaboration on the anomalous couplings of the top quark~\cite{Buckley:2015lku} to EW gauge bosons lead to the bound
\begin{align} \label{eq:dipbound}
\Bigg \lvert \frac{C_{fW}}{C_\text{Yuk}} \Bigg \rvert \lesssim 600 \left( \frac {\Lambda}{10 \TeV} \right)^2  \quad @ \, 95 \% \ \text{C.L.}
\end{align}
Finally, the bound on the anomalous couplings to gluons is
\begin{align} \label{eq:dipbound}
\Bigg \lvert \frac{C_{fG}}{C_\text{Yuk}} \Bigg \rvert \lesssim 100 \left( \frac {\Lambda}{10 \TeV} \right)^2  \quad @ \, 95 \% \ \text{C.L.}
\end{align}
\vskip .5cm

Before moving on to describe the case of the complex TC representation, we conclude the present Section by summarizing some of the main features discussed here: \\ i) Models of fundamental partial compositeness with TC fermions in  real representations of the gauge group require the introduction of two independent complex scalars to give masses  to the top and bottom quarks; ii) As discussed below eq.~\eqref{eq:finetune}, these models require  fine-tuning to accommodate the correct bottom mass and furthermore they suffer from strong constraints coming from the $Zb\bar{b}$ coupling.  

\section{Complex case} \label{sec:complex}
 As template example we consider the  $\SU(N_{TC})$ TC-gauge group featuring fermions in the fundamental representation. Gauge anomalies are avoided by choosing vector like TC-fermions as follows: 
\begin{align}
\psi_\text{TC} \to \  \bigoplus_\alpha \left(\mF + \widetilde \mF\right)_\alpha\,, \quad \alpha = 1, \dots, n_\text{FS} \ ,
\end{align}
with $\mF= (N_\text{TC}, \text{SM})$ and $\widetilde \mF = \overline{({N_\text{TC}}, {\text{SM}})}$.  $\mF$ and $\widetilde \mF$ are both left-handed Weyl fermions, and $\alpha$ is an index that runs over the $n_\text{FS}$ different species. The global fermionic flavour symmetry  is therefore $\SU(N_F)_L \times \SU(N_F)_R$ up to an overall global $\U(1)$ symmetry.  

Using the above conventions, the  relevant Yukawa Lagrangian is conveniently written as  follows\footnote{Again, one should introduce several fermion spurions in the case of multiple scalar species.}
\be \label{eq:yuksun}
\mathcal L_{\text{Yuk}} = (\psi_{L}) ^{i} \phantom{} _a \mathcal F^{a} S_i +  ( \psi_{R}^c) _i \phantom{} ^{ a} \widetilde \mF _{a} S^{*,i}  +\text{h.c.}\,
\ee
Note that, for eq.~\eqref{eq:yuksun} to be a TC singlet, one must assign $S$ and $\mF$ $(\widetilde{\mF})$ respectively to the anti-fundamental and fundamental (anti-fundamental) of $\SU(\ntc)$. 
The transformation properties of $\psi_L$ and $\psi_R ^c$ under the global symmetries are of course different, and they can be summarized as
\begin{align}
\psi_L  & \in \ytableausetup
{aligntableaux=top, boxsize=.7em}  \overline{\begin{ytableau}[]  \\ \end{ytableau}}\, _ \text{L}   \otimes  1_ \text{R} \otimes  \overline{\begin{ytableau}[]  \\ \end{ytableau}}\, _ \text{S}, \quad   \psi_R ^c   \in \ytableausetup
{aligntableaux=top, boxsize=.7em} 1_\text{L} \otimes {\begin{ytableau}[]  \\ \end{ytableau}}\, _ \text{R}   \otimes    \begin{ytableau}[]  \\ \end{ytableau}\, _ \text{S}\,. 
\end{align}
Note that in eq.~\eqref{eq:yuksun} the upper  index $a$ belongs  to the anti-fundamental of $\SU(N_F)_L$, while the lower one to the fundamental of $\SU(N_F)_R$. 

\vspace{.5cm}

The minimal coset with an $\SU(N_{TC})$ gauge group and fermions in the fundamental is $\SU(4)\times \SU(4)$, for which a detailed analysis can be found in Ref.~\cite{Ma:2015gra}.
 Such coset is realized by a fundamental TC-theory with four Dirac fermions that decompose according to the following irreps of $G_\text{SM}$:
\begin{align}
\mF_0 \equiv \left(\ntc, 1, 2, 0\right), \ \mF _\pm \equiv \left({\ntc}, 1, 1, \pm \frac 1 2\right) \,,\nonumber \\
\widetilde{\mF}_0 \equiv \left(\overline \ntc, 1, 2, 0\right), \ \widetilde{\mF} _\pm \equiv \left({\overline\ntc}, 1, 1, \mp \frac 1 2\right) \,.
\end{align}
The above matter content closely resembles that of the most minimal coset, $\SU(4)/\SP(4)$, with the only difference that now we also have the conjugate representation, i.e. all TC-fermions are  Dirac fermions. Because the fundamental irrep of $\gtc$ is now complex, the contraction between TC-scalars and TC-fermions can  be either $\mF \cdot S$ or $\widetilde \mF \cdot S^*$. Therefore, the    fundamental Yukawa Lagrangian for the third family of quarks reads
\begin{align} \label{eq:yuksun1}
\mathcal L = \ &y_Q  Q_\alpha   \mF _{0} ^\alpha   S_q + y_t \,  t_R ^c \widetilde{\mF} _{+}  S_q ^*- y_b \,  b_R^c \widetilde{\mF} _{-}      S_q   ^*\,,
\end{align}
with $S_q  \equiv  \left(\overline 3, 1,-  \frac 1 6\right)$.  Note that there is no way of coupling the field $\widetilde {\mF}_{0}$ to $Q$ -- and similarly  the fields $\mF_{\pm}$ to $(t_R ^c, b_R ^c)$ -- without breaking either Lorentz or SM gauge symmetry. Also, a custodial $\SU(2)_L \times \SU(2)_R$ symmetry is apparent from the explicit expression of the Lagrangian, in the limit where $y_t=y_b$.  


\begin{table}[tb]
\begin{tabular}{ l c c   c | c } 
\hline
 states & $\SU(\ntc)$ &  $\SU(N_F)_L \times \SU(N_F)_R $ & $\SU(N_S)$ & number of states \\ \hline
  $ \widetilde \mF$ &  $\overline \ntc$ & $  1 \, _ \text{L} \times  \Yvcentermath1 \overline {\tiny{ \yng(1)} }\, _ \text{R}$ & 1 & $\ntc \times N_F$   \\ 
  $\mF$ & $\ntc$ & $\Yvcentermath1 \tiny{ \yng(1)}\, _ \text{L} \times 1 \, _ \text{R}$&   1 &  $\ntc \times  N_F $ \\
 $S$  & $\overline \ntc$  & 1 & \tiny{\yng(1)} & $ \ntc   \times 2 N_S  $ \\ \hline
 $S S^* $  & 1 & 1 & $1 \oplus 
\resizebox{2cm}{!}{$
{\ytableausetup
{aligntableaux=center, boxsize=1em}
N_S-1  \left\{ \begin{ytableau}[]  & \\ \\  \\ \none[\vdots]  
 \\ \\ \end{ytableau} \right. }
$}
$  & $ 1\oplus N_S^2-1 $ \\
 $ \widetilde \mF  S^*$ & 1 & $  1_ \text{L}  \times \overline{\tiny{ \yng(1)}} \, _ \text{R} \, $ & $\overline {\tiny{\yng(1)}}$   & $ 2 N_S \times N_F $ \\
 $ \mF S$ & 1 & $ \tiny{ {\yng(1)} \, _ \text{L}} \times 1 \, _ \text{R} $ & $\tiny{\yng(1)}$   & $ 2 N_S \times N_F $ \\
 $\widetilde \mF   \mF $ & 1  & $\tiny{ {\yng(1)}\, _ \text{L}} \times \overline{\tiny{ \yng(1)}} \, _ \text{R}$ & 1 & $ N_F^2$  \\ \hline
\end{tabular}
\caption{Fundamental and composite matter fields for  $\gtc=\SU(\ntc)$.  $N_F$ ($N_S$) is the number of TC-fermions (scalars).}
\label{tab:tab2}
\end{table}

\subsection{Details of the model}
The notation and conventions used here for the $\SU(4)_L\times \SU(4)_R\big / \SU(4)_D$ coset follow those   in Ref.~\cite{Ma:2015gra}. In this paragraph we briefly summarize some of them, which will be useful in the following.

The $\SU(2)_L\times\SU(2)_R$ symmetry is embedded as follows:
\be\label{su4:gen}
T_L ^i = \frac{1}{2}
\left(
\begin{array}{c  c} 
 \mbox{$  \sigma^i $}  & \ 0 \\

 0  \  & \ \mbox{$0$} \  \end{array} \right)\,,
\quad 
T_R ^i = \frac{1}{2}
\left(
\begin{array}{c  c}
\ \mbox{$0$ } & 0 \\
\ \mbox{0}  & \quad \mbox{$ \sigma^i  $ } \end{array} \right)\,.
\ee
The $\SU(4)_L\times \SU(4)_R$ symmetry is made manifest by rearranging the fields in the following four Dirac fermions
\begin{align}
\mF_D^i& = \left( \begin{array}{c} \mF \\ \widetilde \mF^c \end{array} \right) ^i, \quad i =1, \dots, 4\,,  \\
\mF &=  \mF_0 \oplus \mF_\pm \,.
\end{align}
Therefore, the gauge-invariant UV Lagrangian for the free TC-fields reads
\begin{align}
\mathcal L_\text{UV} = i \overline \mF_D^i  \gamma^{\mu} D_\mu \mF_D^i - \mu_L \widetilde \mF_0 \mF_0 - \mu_{R1} \mF_+ \widetilde{\mF}_+ - \mu_{R2} \mF_-  \widetilde{\mF}_{-} \,.
\end{align}
In the following, we will take $\mu_{R1} = \mu_{R2} \equiv \mu_R$, which is the only choice that  preserves the $\SU(2)_R$ symmetry.

We choose the condensate to be aligned along the direction that does not break the EW symmetry, namely 
\begin{align}
\Sigma_0 = \ \left( \begin{array}{c c} \mathbb 1 & 0 \\ 0   & \mathbb 1 \end{array} \right) \,.
\end{align}
The Goldstones transform according to the adjoint representation of the diagonal $\SU(4)_D$ subgroup, in this case the $\mathbf {15}$ dimensional representation, which under the $\SU(2)_L \times \SU(2)_R$ symmetry decomposes as
\begin{align}
\mathbf {15} \to \  H_1(2, 2) + H_2(2,2) + \Delta (3,1) + N (1,3) + s (1,1) \,, 
\end{align}
where we have indicated the name associated with each pNGB multiplet. It is clear that we have two Higgs doublets ($H_1$ and $H_2$) in the spectrum, plus an $\SU(2)_L$ triplet, an $\SU(2)_R$ triplet, and an EW singlet.
For the explicit form of the Goldstone matrix we refer the reader to Ref.~\cite{Ma:2015gra}.

\vskip .2cm

As for the real representation case, we summarize the elementary states and lowest lying scalar and fermionic resonances in Table~\ref{tab:tab2}.

\subsection{Yukawa interactions}
Following the steps of the Section~\ref{sec:real}, we define the spurions associated with the SM quarks 
\begin{align}
\psi_{ L, a} ^i  \ &= \ \left(\begin{array} {cccc} - b_L y_Q  & t_L y_Q   & 0 & 0  \\ 0 &0 &0 &0 \end{array} \right)  \in  \left( 3, 2, - \frac 1 6 \right)\, , \\ 
 (\psi_R ^c ) _i \, ^a  \ &= \ \left(\begin{array} {cccc} 0 & 0 & -y_b  b_R ^c & y_t  t_R ^c  \\ 0 & 0 & 0 & 0 \end{array} \right) \in \left(\overline 3, 1, - \frac 2 3 \right)\oplus \left(\overline 3, 1,  \frac 1 3 \right) \,,
\end{align}
where we have shown the decomposition of each spurion under the SM group, and the subscripts $``L"$ and $``R"$ refer to the fact that they transform in the fundamental of $\SU(4)_L$ and $\SU(4)_R$, respectively.
 Invariants can be easily written down by performing the proper contractions with the matrix $\Sigma$. 

 The  operator constituting the SM Yukawa term is
\begin{align}
\mathcal O _{\text{Yuk}} &= - f \, \left(\psi_{L,a} ^i ( \psi_{R}^c) _{j}\,^b\right) \Sigma^a\phantom{} _b \,\delta^j\phantom{}_i \,.
\end{align}
By expanding the above  term, we find the masses for the top and bottom quarks, together with the linear couplings to the pNGBs:
\begin{align} \label{eq:masscomplex}
\mathcal L_\text{Yuk} \ = \ -  \ C_{\text{Yuk}}  \Bigg\{  \,y_Q  y_t  \, (t_L t_R ^c)  & \left[f \st +i   h_2+\ct (h_1-A_0) -i \st \frac {N_0 + \Delta_0}{\sqrt 2} \right]   \, \nonumber \\
- \  \,y_Q  y_b (b_L b_R ^c)  &  \left[f \st +i  h_2+\ct (  h_1+A_0) +i \st \frac {N_0 + \Delta_0}{\sqrt 2} \right]   \,\nonumber \\
- \ \,y_b  y_Q   (t_L b_R ^c) & \left[ - i \sqrt 2 H^- \ct - i \st \left(N^- + \Delta^-\right) \right]    \nonumber \\
-  \   \, y_t  y_Q (b_L t_R ^c) &  \left[ i \sqrt 2 H^+\ct -i \st \left(N^+ + \Delta^+\right) \right] + \text{h.c.} \Bigg\} \,.
\end{align}
The states $h_1$ and $h_2$ are respectively the real neutral components of $H_1$ and $H_2$, while $A_0$ is the imaginary part of the neutral component of the  doublet $H_1$.

From eq.~\eqref{eq:masscomplex} the top and bottom masses read
\begin{align} \label{eq:qmass}
m_t = \ | C_\text{Yuk} y_Q y_t | v \,, \quad m_b = \ | C_\text{Yuk} y_Q y_b | v\,.
\end{align}
\subsection{Partners}
Differently from the real representation case, the presence of just one TC-scalar implies that no baryons with exotic hypercharges appear in the spectrum.

\subsection{Potential}
The potential generated by gauge boson loops is
\begin{align} \label{eq:vg}
V_{\text{gauge}} &= \ - C_g  f^4 \left( g^2 \Tr\left[T_L ^i \Sigma \, {T_L^i} \Sigma^\dagger\right]+{g'}^2 \Tr \left[T_R ^3 \Sigma \, {T_R^3} \Sigma^\dagger \right] \right) = \nonumber \\
&=  - C_g f^4 \frac{3g^2 + g'^2}{2} \left(\cos^2 \theta  - \frac{\sin \theta \cos \theta}{\sqrt 2} \cdot h_1  \right) + \dots 
\end{align}
while the one generated by the TC-fermion mass term is
\begin{align} \label{eq:vm}
V_m &= \ - C_m f^3 \Tr[M_Q \, \Sigma] + \text{h.c.} = \nonumber \\
 &= - 4\,C_m f^3 (\mu_L +\mu_R)  \cos \theta + \sqrt{2} C_m f^2 (\mu_L+\mu_R) \sin \theta \, h_1 + \dots
\end{align}
where the ellipses stand for terms  carrying higher powers of the pNGB fields.

Concerning the potential generated by the fundamental Yukawa interactions, there are only two operators which are invariant under the full global symmetry (whose diagrammatic representations are given in Fig.~\ref{fig:sundiags})
\begin{align} \label{eq:su4pot} 
O_{V_f}  ^1  &=  \dfrac{f^2 \Lambda^2 }{16 \pi^2}\left(y_{R}^* y_L^* \right)_{a_1} \phantom{} ^{a_2}\phantom{} ^{i_1} \phantom{}_{i_2} \left(y_{L} y_R \right)_{a_3} \phantom{} ^{a_4}\phantom{} ^{i_3} \phantom{}_{i_4}  {\Sigma}^{a_3} \phantom{} _{a_4} 	{\Sigma^\dagger}^{a_1} \phantom{} _{a_2} \delta^{i_2}\phantom{} _{i_1}	\delta^{i_3}\phantom{} _{i_4}\,,   \nonumber \\
O_{V_f} ^2   &=  \dfrac{f^2 \Lambda^2 }{16 \pi^2} \left(y_{R}^* y_L^* \right)_{a_1} \phantom{} ^{a_2}\phantom{} ^{i_1} \phantom{}_{i_2} \left(y_{L} y_R \right)_{a_3} \phantom{} ^{a_4}\phantom{} ^{i_3} \phantom{}_{i_4}  {\Sigma}^{a_3} \phantom{} _{a_4} 	{\Sigma^\dagger}^{a_1} \phantom{} _{a_2}	\delta^{i_4}\phantom{} _{i_1}\delta^{i_2}\phantom{} _{i_3}\,.
\end{align}
The pNGB independent term of the latter two operators is proportional to $\sin^2 \theta$, thus matching the results for heavy quark loops found in \cite{Ma:2015gra}\footnote{In Ref.~\cite{Ma:2015gra} the Yukawa couplings emerge from effective four-fermion interactions that are bilinear in the SM elementary fields.}. Adding up eq.s~(\ref{eq:vg}--\ref{eq:su4pot}), one can rewrite the potential as follows 
\begin{align} \label{pot:su4}
V =& \ (-\widetilde C_g +\widetilde C_t ) \cos 2\theta - \widetilde C_m \, \cos \theta\,,  
\end{align}
where we defined
\begin{align}
\widetilde C_g = & \ C_g f^4 \frac{3g^2 + g'^2}{4}\,, \quad \widetilde C_m = \ C_m f^3 (\mu_L +\mu_R)\,,   \nonumber \\
\widetilde C_t =& \ 3 \, \dfrac{f^2 \Lambda^2 }{32 \pi^2} |y_Q  |^2 \left( |y_b |^2 + |y_t |^2\right)  \left( 3\, C_{V_f}^1 + C_{V_f}^2 \right)\,.
\end{align}
The potential of eq.~\eqref{pot:su4}  is misaligned w.r.t.  $\theta = 0$ for $\widetilde C_t >\widetilde C_g$.
 As a final remark, we notice that the effective potential generated by the partial fundamental Yukawa interactions features only the two operators listed in eq.~\eqref{eq:su4pot}. This has to be contrasted with the three operators appearing in the pseudo-real case~\cite{Cacciapaglia:2017cdi}.  This feature is strictly connected with the  complex versus pseudo-real nature of TC-fermions under the TC-color gauge group.

\vskip .2cm

We checked that the neutral triplet fields, $N_0$ and $\Delta_0$,  do not acquire a VEV because  are  CP-odd. In fact potential tadpoles would have imaginary coefficients. Because the underlying theory is CP even, the operators $O_{V_f} ^{1-2}$ generate potential terms proportional to absolute values of Yukawa couplings, de facto  forbidding tadpole operators.

\tikzset{cross/.style={cross out, draw=black, minimum size=2*(#1-\pgflinewidth), inner sep=0pt, outer sep=0pt},
cross/.default={3pt}}

\begin{figure}
\centering
\begin{align*}
\begin{aligned}
&\begin{tikzpicture}
\begin{feynman}
\vertex (a);
\vertex[above right=1.5 cm of a] (b);
\vertex[below right=1.5cm of b] (c);
\vertex[below = 1.5 cm of c] (d);
\vertex[below left = 1.5 cm of d] (e);
\vertex[above left = 1.5 cm of e] (f);
\diagram[layered layout, small] {
(a) -- [fermion,red,thick] (b) -- [anti fermion,red,thick] (c) -- [ fermion,blue,thick] (d) -- [anti fermion,red,thick] (e) -- [ fermion,red,thick] (f) -- [ anti fermion, blue,thick] (a),
(a) -- [scalar, half left, thick, magenta] (f);
(c) -- [scalar, half right,thick, magenta] (d);
};
\draw[fill=white](1.05,.95) circle(3.mm) node {$\Sigma$};
\draw[fill=white](1.05,-2.4) circle(3.mm) node {$\Sigma^\dagger$};
\end{feynman}
\end{tikzpicture}
&& \hspace{1cm}
\begin{tikzpicture}
\begin{feynman}
\vertex (a);
\vertex[above right=1.5 cm of a] (b);
\vertex[below right=1.5cm of b] (c);
\vertex[below = 1.5 cm of c] (d);
\vertex[below left = 1.5 cm of d] (e);
\vertex[above left = 1.5 cm of e] (f);
\diagram[layered layout, small] {
(a) -- [ fermion,red,thick] (b) -- [anti fermion,red,thick] (c) -- [anti fermion, blue, thick] (d) -- [anti fermion, red, thick] (e) -- [ fermion,red,thick] (f) -- [fermion, blue,thick] (a),
(a) -- [scalar,magenta, thick] (c);
(f) -- [scalar, magenta,thick] (d);
};
\draw[fill=white](1.05,.95) circle(3.mm) node {$\Sigma$};
\draw[fill=white](1.05,-2.4) circle(3.mm) node {$\Sigma^\dagger$};
\end{feynman}
\end{tikzpicture}
\\
&  \hspace{1cm}   O_{V_f} ^1  &&  \hspace{1.8cm} O_{V_f} ^2
\end{aligned}
\end{align*}
\caption{{Diagrams corresponding to the operators $O_{V_f} ^1$ and $O_{V_f}^2$ in eq.~\eqref{eq:su4pot}, contributing to the effective potential of the pNGBs, at the lowest order in the chiral expansion.}}
\label{fig:sundiags}
\end{figure}
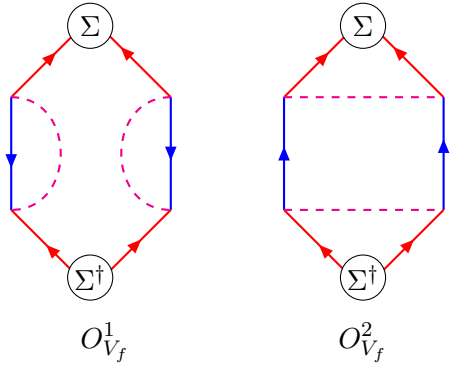

\subsection{Corrections to $Z b_L \bar b_L$}
At  NLO we have two operators contributing to the  $Z b_L \bar b_L$ vertex, one made by left-handed fields and the other made by right-handed ones.   These are: 
\begin{align}
O_{\Pi f} ^L = \ \frac{i f}{\Lambda}\Tr\left[ \psi_{ L} ^\dagger \bar \sigma^\mu \Sigma^\dagger \overleftrightarrow D _\mu \Sigma \psi_{ L} \right]  =& \ \frac g {c_w} |y_Q |^2  \st^2  Z_\mu \left( \bar t_L \gamma^\mu t_L - \bar b_L \gamma^\mu b_L \right)   \nonumber \\
& \qquad +      \frac{g}{\sqrt{2}} |y_Q|^2 \st^2 W_\mu ^+  \overline t_L \gamma^\mu b_L  + \text{h.c.}  \\
O_{\Pi f} ^R = \ \frac{i f}{\Lambda} \Tr\left[ \psi_{ R}  ^\dagger \bar \sigma_\mu \Sigma^\dagger \overleftrightarrow D ^\mu \Sigma \psi_{ R} \right] =& \ \frac g {c_w}  \st^2 Z_\mu \left(|y_t |^2 \bar t_R \gamma^\mu t_R-  |y_b |^2\bar b_R \gamma^\mu b_R \right)  \nonumber \\
& \qquad + \frac{g}{\sqrt 2} y_t^* y_b \st^2 \, W_\mu ^+ \overline t_R \gamma^\mu b_R + \text{h.c.} 
\end{align}
Similarly to eq.~\eqref{eq:zbb}, we get the following constraint on the correction to the $Z b_L \overline b_L $ coupling
\be
C_{\Pi f}^L  |y_Q|^2 \st^2  <   0.05, \quad @ \, 95\% \ \text{CL.}
\ee
This time the above relation can be translated into a bound on $|y_t|$, using eq.~\eqref{eq:qmass}
\begin{align}
\lvert y_t \rvert \frac{ \lvert C_\text{Yuk} \rvert}{\sqrt{\lvert C_{\Pi f} ^L \rvert}} \gtrsim \frac{m_t}{f  \sqrt{0.05}} \approx  \frac{10\TeV}{\Lambda}\,,
\end{align}
which is a reasonable constraint for values of the condensation scale around or above $10 \TeV$.

\subsection{NLO corrections to the kinetic term}
As for the real case, we consider the  NLO  operators that generate corrections to the kinetic term of the non-linear field $\Sigma$. They are listed in Appendix~\ref{sec:NLOsu4}, and the corresponding  contribution to the $\rho$ parameter is
\begin{align}
\delta \rho = \  \frac {3\Lambda^2 \st^2} {4 \pi^2 f^2} \left(3 \,C_{y \Pi D}^1 + C_{y \Pi D}^2  \right)\left(|y_b|^2-|y_t|^2\right) + \frac { \st^2 g'^2} {4 \pi^2 } \left(C_{\Pi D}^1 + C_{\Pi D}^2\right) \,,
\end{align}
where $C_{y\Pi D}^i$ and $C_{\Pi D}^i$ are the strong-dynamics factors multiplying the above-mentioned operators.

 \subsection{Effective interactions for the top sector}
From eq.~\eqref{Alessandro-4tr} we read off the effective four-fermion operator with four right-handed top-quarks
\begin{align}
\mathcal L_\text{EFT} = \ \frac{ C_{4f}^{11} + C_{4f} ^{12}}{2\Lambda} |y_t|^4 (\overline t_R \gamma^\mu t_R)(\overline t_R \gamma_\mu t_R),
\end{align}
leading to the following constraint
\begin{align}
\frac{ C_{4f}^{11} + C_{4f} ^{12}}{2\Lambda} |y_t|^4 < 2.9 \TeV^{-2}  \ \Rightarrow \  | C_{4f}^{11} + C_{4f} ^{12}|^{1/4} |y_t| < 5.8 \left( \frac{\Lambda}{10 \TeV} \right) ^{1/2}, \quad @ \, 95\% \, \text{CL}.
\end{align}

The dipole operators of eq.s~(\ref{eq:dipop1}-\ref{eq:dipop2}) contribute to the couplings between gauge bosons and quarks.
To compute these operators, we recall that the EW and color generators are embedded in the $\SU(4)_D$ and $\U(3)_S$ flavour symmetries respectively as follows
\renewcommand{\arraystretch}{0.9}
\begin{align}
A_\mu ^I T_{F} ^I = \ \frac 1 2  
\left( \begin{array}{cc } g W_\mu ^i \sigma^i   & 0  \\
0 & g' \sigma^3 B_\mu  \end{array} \right) \,,    \quad 
G_\mu ^A T_\mathcal{S} ^A = \  \frac  {g_s} 2 G_\mu ^a \lambda_a 
-  \frac {g'} 6  B_\mu 
 \mathbb 1 _3 \,.
\end{align}

The dipole operators of eq.s~(\ref{eq:dipop1}-\ref{eq:dipop2}) generate the following couplings between gauge bosons and quarks:
\begin{align} 
\mathcal O _{fW} &= \ -\frac{m_t}{8   C_\text{Yuk} \Lambda^2 v} (1+\ct) \, g \mathcal O_{uW}^{33 *}  + \ \dots	  \\
\mathcal O _{fG} &= \  \frac{m_t}{2 C_\text{Yuk} \Lambda^2 v}  (1+\ct) \, \left( g_s \mathcal O _{uG}^{33 *} - \frac {1} {6} g' \mathcal O_{uB}^{33 *}  \right) + \ \dots
\end{align}
As before, comparing with the results of the \texttt{TopFitter} collaboration, we get  bounds on the strong dynamics coefficients factors. From the couplings to EW gauge bosons we have
\begin{align} \label{dipboundcompl1}
\Bigg \lvert \frac{C_{fW}}{C_\text{Yuk}} \Bigg \rvert \lesssim 2300 \left( \frac {\Lambda}{10 \TeV} \right)^2  \quad @ \, 95 \% \ \text{C.L.}
\end{align}
while from the anomalous couplings to gluons we end up with
\begin{align} \label{dipboundcompl2}
\Bigg \lvert \frac{C_{fG}}{C_\text{Yuk}} \Bigg \rvert \lesssim 200 \left( \frac {\Lambda}{10 \TeV} \right)^2  \quad @ \, 95 \% \ \text{C.L.}
\end{align}
\section{Conclusions}

We are now in a position to offer our conclusions, by summarizing the main results. The analysis performed here aimed at exhausting the effective field theories at the electroweak scale for  minimal models of  fundamental partial compositeness. Since the case with TC-fermions in the pseudo-real representation was  considered in \cite{Cacciapaglia:2017cdi} we took here the TC-fermions either in the real or complex representation of the gauge group underlying the composite Higgs dynamics. Due to the different TC-fermion nature, the cosets  are respectively  SU(5)/SO(5) and  SU(4)$\times$SU(4)/SU(4)$_D$. The electroweak theory is embedded into the maximal diagonal subgroups i.e. SO(5) and SU(4)$_D$. The TC-fermion nature plays a crucial role when trying to construct the partial fundamental mass operators for the SM fermions. This is so since the TC-fermions, by construction, do not carry QCD color. The latter is carried by new TC-scalars whose quantum numbers with respect to the SM must  yield renormalizable TC and SM gauge singlet operators involving one SM fermion, one TC fermion and a TC-scalar. The TC-scalar spectrum, de facto, reflects the nature of the TC-fermion sector with important consequences for the low energy predictions for the SM fermion Yukawa structure as well as the structure of   higher order operators including the interactions with SM gauge bosons and TC pNGBs. Given the above we provided in the appendices a complete list of the effective operators emerging at the electroweak scale stemming from the elementary theory.  

To determine the viability of the theories we studied the vacuum alignment and stability of the vacuum against condensation induced by possible tad-pole interactions, the electroweak precision constraints and further collider constraints. We also investigated mass generation for the third family of quarks. Last but not  least we discussed the main differences among the different models of minimal partial compositeness. For example, for the pseudo-real and complex representation  the top and bottom mass difference comes from splitting the right-handed fundamental Yukawa couplings. However, for the real representation case such a difference cannot be attributed only to the right-handed fundamental Yukawa couplings since these theories feature two distinct left-handed partial composite Yukawa interactions involving two TC-scalars. 

Future directions include an in-depth study of the composite massive and massless spectrum stemming from the new fundamental underlying dynamics  via first principle lattice simulations, along the lines of  
\cite{Hansen:2017mrt}.  We plan a more general investigation of the spectrum of the theory using the effective Lagrangian as function of the effective parameters. A detailed study of collider phenomenology including decays of composite fermions such as top-partners will  be explored as well. Furthermore, in these theories it is by no mean obvious that the first massive states to discover at present and future colliders will be spin one resonances. In fact spin zero composite states made by TC-scalars are expected to be observed. It would therefore be interesting to explore the discovery potential at future colliders of spin zero composite states made by two TC-scalars, some of which will have lepto-quarks quantum numbers \cite{Blum:2016szr}. Finally, flavour dynamics and modeling is highly interesting and of immediate impact for the investigation of flavour observables and anomalies.  This requires, as done for the pseudo-real case, to include at the fundamental and effective level the operators describing light generations of quarks and leptons.  

\acknowledgments{We thank Anders Eller Thomsen for relevant discussions.  The authors are partially supported by the Danish National Research Foundation grant DNRF:90.}

\newpage

\appendix  \label{sec:app}

\section{Real case}
In this appendix we provide all the missing operators that have been used in Section~\ref{sec:real}, in the case of a real TC representation.

\subsection{List of four-fermion operators}
In this appendix we list all the four-fermion operators that are generated at NLO in the effective theory.

The self-hermitian operators are the following five:
	\begin{align} \label{eq:fourferm1}
	\mathcal{O}_{4f}^1 &= \dfrac{1}{4 \Lambda^2} (\spur{i_1}{a_1} \spur{i_2}{a_2} ) (\spurbar{i_3}{a_3} \spurbar{i_4}{a_4} ) \Sigma^{a_1 a_2} \Sigma^\dagger_{a_3 a_4} \omega_{i_1 i_2} \omega_{i_3 i_4}\ , \\
	\mathcal{O}_{4f}^2 &= \dfrac{1}{4 \Lambda^2} (\spur{i_1}{a_1} \spur{i_2}{a_2} ) (\spurbar{i_3}{a_3} \spurbar{i_4}{a_4} ) \left(\delta^{a_1}_{\enspace a_3} \delta^{a_2}_{\enspace a_4} + \delta^{a_1}_{\enspace a_4} \delta^{a_2}_{\enspace a_3} \right) \omega_{i_1 i_2} \omega_{i_3 i_4}\ , \\
	\mathcal{O}_{4f}^3 &= \dfrac{1}{4 \Lambda^2} (\spur{i_1}{a_1} \spur{i_2}{a_2} ) (\spurbar{i_3}{a_3} \spurbar{i_4}{a_4} ) \Sigma^{a_1 a_2} \Sigma^\dagger_{a_3 a_4} \left(\omega_{ i_1 i_4} \omega_{ i_2 i_3} + \omega_{ i_1 i_3} \omega_{ i_2 i_4} \right)\ ,	\\
	\mathcal{O}_{4f}^4 &= \dfrac{1}{4 \Lambda^2} (\spur{i_1}{a_1} \spur{i_2}{a_2} ) (\spurbar{i_3}{a_3} \spurbar{i_4}{a_4} ) \left( \delta^{a_1}_{\enspace a_3} \delta^{a_2}_{\enspace a_4} \omega_{ i_1 i_3} \omega_{ i_2 i_4} + \delta^{a_1}_{\enspace a_4} \delta^{a_2}_{\enspace a_3} \omega_{ i_1 i_4} \omega_{ i_2 i_3}\right)\ , \\
	\mathcal{O}_{4f}^5 &= \dfrac{1}{4 \Lambda^2} (\spur{i_1}{a_1} \spur{i_2}{a_2} ) (\spurbar{i_3}{a_3} \spurbar{i_4}{a_4} ) \left( \delta^{a_1}_{\enspace a_3} \delta^{a_2}_{\enspace a_4} \omega_{ i_1 i_4} \omega_{ i_2 i_3} + \delta^{a_1}_{\enspace a_4} \delta^{a_2}_{\enspace a_3} \omega_{ i_1 i_3} \omega_{ i_2 i_4}\right)\ , \label{eq:fourfermion5}
	\end{align}
where we have defined $\bar \psi^{i a} \equiv \bar \psi _j  \phantom{} ^{a}   \omega^{ij}$. Additionally we have the following three non-hermitian operators:
\begin{align}
	\mathcal{O}_{4f}^6 &= \dfrac{1}{8 \Lambda^2} (\spur{i_1}{a_1} \spur{i_2}{a_2} ) (\spur{i_3}{a_3} \spur{i_4}{a_4} )  \Sigma^{a_1 a_2} \Sigma^{a_3 a_4} \omega_{i_1 i_2} \omega_{i_3 i_4}\,, \label{eq:fourfermion6} \\
	\mathcal{O}_{4f}^7 &= \dfrac{1}{8 \Lambda^2} (\spur{i_1}{a_1} \spur{i_2}{a_2} ) (\spur{i_3}{a_3} \spur{i_4}{a_4} )  \left(\Sigma^{a_1 a_4} \Sigma^{a_2 a_3} + \Sigma^{a_1 a_3} \Sigma^{a_2 a_4}\right) \omega_{i_1 i_2} \omega_{i_3 i_4}\,,   \\
	\label{eq:fourferm8}
	\mathcal{O}_{4f}^8 &= \dfrac{1}{8 \Lambda^2} (\spur{i_1}{a_1} \spur{i_2}{a_2} ) (\spur{i_3}{a_3} \spur{i_4}{a_4} )  \Sigma^{a_1 a_2} \Sigma^{a_3 a_4} \left(\omega_{i_1 i_4} \omega_{i_2 i_3} + \omega_{i_1 i_3} \omega_{i_2 i_4}\right)\,.
	\end{align}	
After expanding the above operators, we reorganize them according to the following six classes:

\vskip .3cm

-- Operators with four left-handed quarks:

\noindent \begin{align} \label{4f:1}
\mathcal L _\text{EFT} \supset \ \  \frac{ \st^4 C_{4f}^3 |y_Q|^4 + \left(|y_Q|^4+|\tilde y_Q|^4\right) (C_{4f}^4 + C_{4f}^5) } {4\Lambda^2}  &(\tLgtL)(\tLgtL)  \ \nonumber \\
+ \  \frac 1 2 \frac{\ct^4 |y_Q \tilde y_Q|^2 C_{4f}^3 + 2 \left(|y_Q|^2+|\tilde y_Q|^2\right)^2 C_{4f}^4}{4\Lambda^2  } &(\bLgbL)(\tLgtL) \  \nonumber \\
 \frac 1 2 \frac{-\ct^2 |y_Q \tilde y_Q|^2 C_{4f}^3 + 2 \left(|y_Q|^4+|\tilde y_Q|^4\right) C_{4f}^5}{4\Lambda^2} & (\bLgtL)(\tLgbL) \nonumber \\
+\  &   (t \leftrightarrow b, \ y_Q \leftrightarrow \tilde y_Q )\,.
\end{align} 
-- Operators with four right-handed quarks:
\noindent \begin{align}
\label{Alessandro-4tr}
\mathcal L _\text{EFT} \supset \ |y_t|^4 \frac{\ctt^2 C_{4f}^3 + C_{4f}^4 +C_{4f}^5}{4\Lambda^2} & (\tRgtR)(\tRgtR)  \nonumber \\
\frac {|y_t y_b|^2} 2 \frac{\ctt^2 C_{4f}^3 + 2 C_{4f}^4 + 2 C_{4f}^5}{4\Lambda^2}&(\bRgbR)(\tRgtR) \nonumber \\
+\  &   (t \leftrightarrow b, \ y_t \leftrightarrow \tilde y_b )\,.
\end{align}
-- Operators with two left-handed and two right-handed top quarks:
\begin{align}
\mathcal L_\text{EFT} \supset \ |y_t y_Q|^2 \frac{ \stt^2 C_{4f}^1+ C_{4f}^2}{2\Lambda^2}(\bar t_R t_L)(\bar t_L t_R) + \frac{ \stt^2 C_{4f}^3 + 2C_{4f}^4}{8\Lambda^2}& (\tLgtL)(\tRgtR) \nonumber \\
+\  (y_Q y_t)^2 \st^2 \frac{-2 \ct^2 C_{4f}^6 -  C_{4f}^7 (\ctt + \ct)^2 +   C_{4f}^8 (\frac 2 3 \ctt - \ct^2)}{\Lambda^2}&(\bar t_R t_L)(\bar t_R t_L) \nonumber \\
+ \ (y_Q y_t)^2 \st^2 \ctt \frac{C_{4f}^8}{\Lambda^2} \  &(\bar t_R T^A t_L)(\bar t_R T^A t_L)\,.
\end{align}
-- Operators with two left-handed and two right-handed bottom quarks:
\begin{align}
\mathcal L_\text{EFT} \supset \ |y_b y_Q|^2 \frac{ \stt^2 C_{4f}^1+ C_{4f}^2}{2\Lambda^2}(\bar b_R b_L)(\bar b_L b_R) + \frac{ \stt^2 C_{4f}^3 + 2C_{4f}^4}{8\Lambda^2}& (\bLgbL)(\bRgbR) \nonumber \\
+\  (y_Q y_b)^2 \st^2 \frac{-2 \ct^2 C_{4f}^6 -  C_{4f}^7 (\ctt + \ct)^2 +   C_{4f}^8 (\frac 2 3 \ctt - \ct^2)}{\Lambda^2}&(\bar b_R b_L)(\bar b_R b_L) \nonumber \\
+ \ (y_Q y_b)^2 \st^2 \ctt \frac{C_{4f}^8}{\Lambda^2} \  &(\bar b_R T^A b_L)(\bar t_R T^A t_L)\,.
\end{align}
-- Operators with two left-handed bottom and two right-handed top quarks, or vice versa:
\begin{align}
\mathcal L_\text{EFT} \supset \ &   \frac{ C_{4f}^2}{2 \Lambda^2} \left[|y_b \tilde y_Q|^2 (\bar t_R b_L)(\bar b_L t_R) + |y_t y_Q|^2 (\bar b_R t_L)(\bar t_L t_R)  \right] \nonumber \\
+ \ & |y_t|^2 \frac{\st^2 \ct^2 |\tilde y_Q|^2 C_{4f}^3 +\left(|y_Q|^2+|\tilde y_Q|^2\right)C_{4f}^4}{2\Lambda^2}(\bLgbL)(\tRgtR) \nonumber \\
+ \ & |y_b|^2 \frac{\st^2 \ct^2 | y_Q|^2 C_{4f}^3 +\left(|y_Q|^2+|\tilde y_Q|^2\right)C_{4f}^4}{2\Lambda^2}(\tLgtL)(\bRgbR)\,.
\end{align}
-- Operators with a left-handed and right-handed top quark and a left-handed and right-handed bottom quark:
\begin{align}\label{4f:6}
\mathcal L_\text{EFT} \supset \ & y_t y_Q (\tilde y_Q y_b)^*  \stt^2 \frac{C_{4f}^1}{2\Lambda^2} \left[ (\bar t_R t_L)(\bar b_L b_R)   (\bar b_R b_L)(\bar t_L t_R) \right]  \nonumber \\
- \ & y_b \tilde y_Q (y_t y_Q)^* \stt^2 \frac{C_{4f}^8}{\Lambda^2} (\bar b_R T^A b_L)(\bar t_R T^A t_L) \nonumber \\
+ \ &  y_Q \tilde y_Q y_t y_b \frac{ -2 \ctt C_{4f}^7 + \frac 4 3 \st^2 (\ctt - 2\ct^2) C_{4f}^8}{\Lambda^2} (\bar b_R t_L)(\bar t_R b_L)  \nonumber  \\
+ \ & y_Q \tilde y_Q y_t y_b \frac{-  \stt^2 C_{4f}^6 - 2 C_{4f}^7(\ctt+\ct^2\st^2)+\frac 1 3 \st^2 C_{4f}^8(\ctt-2\ct^2)}{\Lambda^2} (\bar b_R b_L)(\bar t_R t_L) 
\end{align}

\subsection{Dipole operators}

Dipole operators appear at NLO, contributing to the couplings among SM fermions and EW and QCD gauge bosons. These operators  have been used to obtain the collider constraint of eq.~\eqref{eq:dipbound}, and they read: 
\begin{align} \label{eq:dip1} 
\mathcal O _{fW} = \  \frac{ f}{2 \Lambda^2} (\spur{i_1}{a_1} \sigma^{\mu \nu} \spur{i_2}{a_2}) A_{\mu \nu} ^I \left(T^I  _{ F} \Sigma + \Sigma (T^I _{ F})^T  \right) ^{a_1 a_2} \omega_{i_1 i_2} \,,
\end{align}
\begin{align} \label{eq:dip2}
\mathcal O _{fG} = \  \frac{ f}{2 \Lambda^2} (\spur{i_1}{a_1} \sigma^{\mu \nu} \spur{i_2}{a_2}) G_{\mu \nu} ^A \Sigma^{a_1 a_2} \left(\omega T^A _S + (T^A _S)^T  \omega \right) _{i_1 i_2}\,,
\end{align}
where $T_{ F} ^I$ and $T_{ S}^A$  contain the generators   of the  EW and QCD color subgroups of the unbroken flavour symmetries, with field strengths $A_{\mu \nu}^I$ and $G_{\mu \nu}^A$, respectively.

\subsection{List of NLO kinetic operators}\label{app:NLOSU5}
Here we provide all the operators with two covariant derivatives and four insertions of the Yukawa spurion $y_f$.
\begin{align} \label{4f:first}
\mathcal{O}_{y\Pi D}^{1-3} &= \dfrac{1}{4} \dfrac{\Lambda^2}{16 \pi^2} (y_f^\ast y_f)^{a_1} \phantom{}_{a_2} \phantom{}^{i_1}\phantom{}^ {i_2} (y_{f'}^\ast y_{f'})^{a_3} \phantom{}_{a_4} \phantom{}^{i_3}\phantom{}^ {i_4}  \SDS{1}{2} \SDS{3}{4}\omega_{i_1}\phantom{}_{ i_2} \omega_{i_3}\phantom{}_{ i_4}\, ,\\
\mathcal{O}_{y\Pi D}^{4-6} &= \dfrac{1}{4} \dfrac{\Lambda^2}{16 \pi^2} (y_f^\ast y_f)^{a_1} \phantom{}_{a_2} \phantom{}^{i_1}\phantom{}^ {i_2} (y_{f'}^\ast y_{f'})^{a_3} \phantom{}_{a_4} \phantom{}^{i_3}\phantom{}^ {i_4} \SDS{1}{4} \SDS{3}{2}\omega_{i_1}\phantom{}_{ i_2} \omega_{i_3}\phantom{}_{ i_4}\, , \\
\mathcal{O}_{y\Pi D}^{7-9} &= \dfrac{\Lambda^2}{16 \pi^2}(y_f^\ast y_f)^{a_1} \phantom{}_{a_2} \phantom{}^{i_1}\phantom{}^ {i_2} (y_{f'}^\ast y_{f'})^{a_3} \phantom{}_{a_4} \phantom{}^{i_3}\phantom{}^ {i_4} \DSd{1}{3}\DS{2}{4} \omega_{i_1}\phantom{}_{ i_2} \omega_{i_3}\phantom{}_{ i_4}\, ,  \\
\label{eq:NLOcorr4}
\mathcal{O}_{y\Pi D}^{10-12} &= \dfrac{\Lambda^2}{16 \pi^2}(y_f^\ast y_f)^{a_1} \phantom{}_{a_1} \phantom{}^{i_1}\phantom{}^ {i_2} (y_{f'}^\ast y_{f'})^{a_3} \phantom{}_{a_4} \phantom{}^{i_3}\phantom{}^ {i_4} \DSd{3}{5}\DS{5}{4}  \omega_{i_1}\phantom{}_{ i_2} \omega_{i_3}\phantom{}_{ i_4}\, ,\\
\label{eq:NLOcorr5}
\mathcal{O}_{y\Pi D}^{13-15} &= \dfrac{\Lambda^2}{16 \pi^2}(y_f^\ast y_f)^{a_1} \phantom{}_{a_2} \phantom{}^{i_1}\phantom{}^ {i_2} (y_{f'}^\ast y_{f'})^{a_3} \phantom{}_{a_4} \phantom{}^{i_3}\phantom{}^ {i_4} \DSd{1}{5}\DS{5}{3} \omega_{i_1}\phantom{}_{ i_2} \omega_{i_3}\phantom{}_{ i_4}\, ,\\
\label{4f:last}
\mathcal{O}_{y\Pi D}^{16-18} &= \dfrac{1}{2} \dfrac{\Lambda^2}{16 \pi^2} (y_f^\ast y_f)^{a_1} \phantom{}_{a_2} \phantom{}^{i_1}\phantom{}^ {i_2} (y_{f'}^\ast y_{f'})^{a_3} \phantom{}_{a_4} \phantom{}^{i_3}\phantom{}^ {i_4}  \Sigma^\dagger_{a_1a_3} \DS{2}{5} \SDS{5}{4}\omega_{i_1}\phantom{}_{ i_2} \omega_{i_3}\phantom{}_{ i_4}\, .
\end{align}
For each of the above templates we indicated only one possible contraction of the scalar $\SO(2N_S)$ symmetry indices, leaving understood that the two additional contractions shown in eq.s~(\ref{eq:OVf1}--\ref{eq:OVf3}) are also possible.

One must also consider corrections stemming from the propagation of one SM gauge boson,  described by the following two operators
\begin{align} 
O_{\Pi D} ^1&= \frac{1}{4}\frac{f^2}{16\pi^2} g_i^2\Tr\left[\left(\Sigma \overleftrightarrow D_\mu \Sigma^\dagger\right)T^i \left(\Sigma \overleftrightarrow D_\mu \Sigma^\dagger\right)T^i\right] \,, 
\nonumber \\
\label{eq:NLOgauge}
O_{\Pi D} ^2 &=  \frac{1}{4}\frac{f^2}{16\pi^2} g_i ^ 2\Tr\left[\left(\Sigma \overleftrightarrow D_\mu \Sigma^\dagger\right)T^i\right] \, \Tr\left[\left(\Sigma \overleftrightarrow D^\mu \Sigma^\dagger\right)T^i\right]\,.
\end{align} 
Some of the kinetic operators are found to give corrections to the $\rho$ parameter that do not vanish as $\st \to 0$. Such operators can be however reabsorbed by a unitary transformation of the non-linear field $\Sigma$, their effects being therefore non-physical. The contribution of the remaining operators is proportional to $\st^2$, and is summarized in eq.~\eqref{eq:deltarho}, where the coefficients $C_{y \Pi D}$ and $C_{ \Pi D}$ take the following values:
\begin{align} \label{eq:cypd}
C_{y \Pi D} &=  \  3 \, C_{y\Pi D} ^1+  C_{y\Pi D} ^3 +  \frac   {3 C_{y\Pi D} ^4}{2} + \frac   { C_{y\Pi D} ^6}{2} - 3 \,C_{y\Pi D} ^7 -  C_{y\Pi D} ^9 - 3 C_{y\Pi D} ^{16} - \frac{C_{y\Pi D} ^{17} }{2} \, , \nonumber \\
C_{ \Pi D} &= \  \frac {C_{\Pi D}^1}{2} + C_{\Pi D}^2\,.
\end{align}

\section{Complex case}
In this appendix we provide all the missing operators that have been used in Section~\ref{sec:complex}, in the case of a complex TC representation.

\subsection{List of four-fermion operators}
Here we list all the four-fermion operators that are generated at NLO in the effective theory.
\vskip .2cm

There are four complex  
\begin{align}
\mathcal O_1 = \dfrac{1}{8 \Lambda^2}  (\psi_{L,a} ^i   \psi_{L, b } ^j ) (\psi_{R,i} ^{a'}   \psi_{R, j } ^{b'} ) \Sigma^a \phantom{} _{a'} \Sigma^b \phantom{} _{b'}\,, \\
\mathcal O_2 = \dfrac{1}{8 \Lambda^2}  (\psi_{L,a} ^i   \psi_{L, b } ^j ) (\psi_{R,j} ^{a'}   \psi_{R, i } ^{b'} ) \Sigma^a \phantom{} _{a'} \Sigma^b \phantom{} _{b'}\,,\\
\mathcal O_3 = \dfrac{1}{8 \Lambda^2}  (\psi_{L,a} ^i \psi_{R,i} ^{a'} )  (\psi_{L,b} ^j \psi_{R,j} ^{b'} ) \Sigma^a \phantom{} _{a'} \Sigma^b \phantom{} _{b'}\,, \\
\mathcal O_4 = \dfrac{1}{8 \Lambda^2}  (\psi_{L,a} ^i \psi_{R,j} ^{a'} )  (\psi_{L,b} ^j \psi_{R,i} ^{b'} ) \Sigma^a \phantom{} _{a'} \Sigma^b \phantom{} _{b'} \,,
\end{align}
and eight  self-hermitian  operators
\begin{align}
\mathcal O_5 & = \dfrac{1}{4 \Lambda^2}   (\psi_{L,a} ^i   \psi_{L, b } ^j )  (\bar \psi_{L,i} ^{a}  \bar  \psi_{L, j } ^{b} ) \,,\\
\mathcal O_6 &= \dfrac{1}{4 \Lambda^2}   (\psi_{L,a} ^i   \psi_{L, b } ^j )  (\bar \psi_{L,j} ^{a}  \bar  \psi_{L, i } ^{b} )\,, \\
\mathcal O_7 &=\dfrac{1}{4 \Lambda^2}   (\psi_{L,a} ^i    \psi_{R,i} ^{b} )  (\bar \psi_{L,j} ^{a}  \bar \psi_{R, b } ^j )\,, \\
\mathcal O_8 &=\dfrac{1}{4 \Lambda^2}   (\psi_{L,a} ^i    \psi_{R,j} ^{b} )  (\bar \psi_{L,i} ^{a}  \bar \psi_{R, b } ^j ) \,,\\
\mathcal O_9 &=\dfrac{1}{4 \Lambda^2}   (\psi_{L,a} ^i    \psi_{R,i} ^{a'} )  (\bar \psi_{L,j} ^{b}  \bar \psi_{R, b' } ^j ) \Sigma^a \phantom{} _{a'} {\Sigma^\dagger}^{b'} \phantom{} _{b}\, ,\\
\mathcal O_{10} &= \dfrac{1}{4 \Lambda^2}   (\psi_{L,a} ^i    \psi_{R,j} ^{a'} )  (\bar \psi_{L,i} ^{b}  \bar \psi_{R, b' } ^j ) \Sigma^a \phantom{} _{a'} {\Sigma^\dagger}^{b'} \phantom{} _{b}\,, \\
\mathcal O_{11} &= \dfrac{1}{4 \Lambda^2}  (\psi_{R,i} ^{a}   \psi_{R, j } ^{b} ) (\bar \psi_{R,a} ^i  \bar \psi_{R, b } ^j ) \,,\\
\mathcal O_{12} &= \dfrac{1}{4 \Lambda^2}  (\psi_{R,i} ^{a}   \psi_{R, j } ^{b} ) (\bar \psi_{R,a} ^j  \bar \psi_{R, b } ^i ) \, ,
\end{align}
where as usual  Lorentz indices are contracted inside brackets.  After expanding the above operators, we reorganize them according to the following six classes:

%

\vskip .3cm

-- Operators with four left-handed quarks:

\noindent \begin{align}
\mathcal L _\text{EFT} \supset \  \frac{ C_{4f}^5+ C_{4f}^6}{2} \frac{|y_Q|^4}{\Lambda^2} \Bigg[(\overline t_L \gamma^\mu t_L) (\overline t_L \gamma_\mu t_L)+ (\overline b_L \gamma^\mu b_L)(\overline b_L \gamma_\mu b_L)+2 (\overline t_L \gamma^\mu t_L)(\overline b_L \gamma_\mu b_L)\Bigg]\,.
\end{align}

-- Operators with four right-handed quarks:
\noindent \begin{align}
\label{Alessandro-4tr}
\mathcal L _\text{EFT} \supset \  & \frac{ C_{4f}^{11}+ C_{4f}^{12}}{2} \frac{ |y_t|^4}{\Lambda^2}  (\overline t_R \gamma^\mu t_R) (\overline t_R \gamma_\mu t_R)+ \frac{ C_{4f}^{11}+ C_{4f}^{12}}{2} \frac{ |y_b|^4}{\Lambda^2} (\overline b_R \gamma^\mu b_R)(\overline b_R \gamma_\mu b_R) \nonumber \\ 
     &  \qquad + \   (C_{4f}^{11}+ C_{4f}^{12})\frac{ |y_t y_b|^2}{\Lambda^2} (\overline t_R \gamma^\mu t_R)(\overline b_R \gamma_\mu b_R) \,.
\end{align}

-- Operators with two left-handed and two right-handed top quarks:
\noindent \begin{align}
\mathcal L _\text{EFT} \supset \  \frac{|y_Q y_t|^2}{\Lambda^2}  & \Bigg[\frac 1 2 \left(\frac 1 3 C_{4f}^7 + C_{4f}^8 -C_{4f}^{10} \st \right) (\overline t_L \gamma^\mu t_L) (\overline t_R \gamma_\mu t_R)    \nonumber \\
&  \qquad - \st C_{4f}^9(\overline t_L t_R) (\overline t_R t_L)    +   C_{4f}^7 (\overline t_L T^A \gamma^\mu t_L) (\overline t_R T^A \gamma_\mu t_R)	\Bigg]  \nonumber \\
+ \  \frac{ y_Q^2 y_t^2}{\Lambda^2}  \st^2 \, & \Bigg[ \frac{-4 C_{4f}^1 - 4 C_{4f}^2 + 3 C_{4f}^3 +  C_{4f}^4}{3} (\overline t_R t_L)(\overline t_R t_L)   \nonumber \\
&  \qquad  + 2(-C_{4f}^1 - C_{4f}^2 + C_{4f}^4)  (\overline t_R T^A t_L)(\overline t_R T^A t_L) \Bigg]\,.
\end{align}

-- Operators two left-handed and two right-handed bottom quarks:
\noindent 
\begin{align}
\mathcal L _\text{EFT} \supset \ \frac{|y_Q y_b|^2}{\Lambda^2}  & \Bigg[\frac 1 2 \left(\frac 1 3 C_{4f}^7 + C_{4f}^8 -C_{4f}^{10} \st \right) (\overline b_L \gamma^\mu b_L) (\overline b_R \gamma_\mu b_R)   \nonumber \\
&    \qquad - \st C_{4f}^9(\overline b_L b_R) (\overline b_R b_L) + C_{4f}^7 (\overline b_L T^A \gamma^\mu b_L) (\overline b_R T^A \gamma_\mu b_R)	\Bigg] \nonumber \\
 +  \frac{ y_Q^2 y_b^2}{\Lambda^2}  \st^2 \, & \Bigg[ \frac{-4 C_{4f}^1 - 4 C_{4f}^2 + 3 C_{4f}^3 +  C_{4f}^4}{3} (\overline b_R b_L)(\overline b_R b_L)   \nonumber \\
&  \qquad + 2(-C_{4f}^1 - C_{4f}^2 + C_{4f}^4)  (\overline b_R T^A b_L)(\overline b_R T^A b_L) \Bigg]\,.
\end{align}

-- Operators with two left-handed bottom and two right-handed top quarks, or vice versa:
\begin{align}
\mathcal L _\text{EFT} \supset \ \frac 1 2 C_{4f}^8 \frac{|y_Q |^2}{\Lambda^2} \Bigg[|y_b|^2 (\overline t_L \gamma^\mu t_L) (\overline b_R \gamma^\mu b_R)+|y_t|^2 (\overline b_L \gamma^\mu b_L) (\overline t_R \gamma^\mu t_R)  \Bigg]\,.
\end{align}

-- Operators with a left-handed and right-handed top quark and a left-handed and right-handed bottom quark:
\begin{align} 
\mathcal L _\text{EFT} \supset \ & \frac{|y_Q |^2}{\Lambda^2}  \st^2 \left(-\frac 4 3 C_{4f}^1 - \frac 1 3 C_{4f}^2 + C_{4f}^3+\frac 1 3 C_{4f}^4\right)\Bigg[ y_t y_b^* (\overline t_R t_L)(\overline b_L b_R)	+y_t^* y_b (\overline b_R b_L)(\overline t_L t_R)	\Bigg] \nonumber \\
 - \  &     \frac 1 2 \frac{|y_Q |^2}{\Lambda^2}  \st  \, C_{4f}^{10} \Bigg[y_t y_b^* (\overline t_L \gamma^\mu b_L)(\overline b_R \gamma_\mu t_R)+ y_b y_t^*  (\overline b_L \gamma^\mu t_L)(\overline t_R \gamma_\mu b_R) \Bigg] \nonumber \\
+ &  \  \frac{y_Q ^2 y_t y_b}{\Lambda^2}\st^2  \Bigg[ -2 C_{4f}^1 (\overline b_R T^A t_L)(\overline t_R T^A b_L)+ 2(-C_{4f}^2 + C_{4f}^4) (\overline b_R T^A b_L)(\overline t_R t^A t_L)  \nonumber \\
& \hspace{2cm}  + \ \left(-C_{4f}^1 - \frac 1 3 C_{4f}^2 + \frac 1 3 C_{4f}^3 +\frac 1 3 C_{4f}^4 \right) (\overline b_L b_R)(\overline t_R t_L)  \nonumber  \\
& \hspace{6cm} + \left(  - \frac 1 3 C_{4f}^1  - C_{4f}^2\right) (\overline t_R b_L)(\overline b_R t_L) + \text{h.c.} \Bigg]
\end{align}

\subsection{Dipole operators} 

As for the real case here as well dipole operators appear at NLO, contributing to the couplings among SM fermions and gauge bosons. These operators  have been used to obtain the collider constraint in eq.~(\ref{dipboundcompl1}--\ref{dipboundcompl2}), and they read 
\begin{align} \label{eq:dipop1} 
\mathcal O _{fW} =& \  \frac{ f}{2 \Lambda^2} (\psi_{ L, a} ^i   \sigma^{\mu \nu} (\psi_{R}^c) _{i}\,^b) A_{\mu \nu} ^I \left(T^I  _{ F} \,\Sigma\right)^{a}\, _b  \,, \\ \nonumber \\
 \label{eq:dipop2}
\mathcal O _{fG} =& \  \frac{ f}{2 \Lambda^2} (\psi_{ L, a} ^i   \sigma^{\mu \nu} (\psi_{R}^c) _{j}\,^b) G_{\mu \nu} ^A \Sigma^a \,_b( {T^A_S}) _{i} \,^{j}\,,
\end{align}
where $T_{ F}$ and $T_{ S}$  contain the generators  of the fermionic $\SU(N_F)_D$ and scalar $\U(N_S)$  symmetries, and the field strengths $A_{\mu \nu}$ and $G_{\mu \nu}$ are associated with the  gauged EW and QCD color subgroups.

\subsection{Corrections to the kinetic term}\label{sec:NLOsu4}
Here we provide all the operators with two covariant derivatives and four insertions of the Yukawa spurions.
At NLO, there are five different contractions among the $\SU(N_F)\times\SU(N_F)$ indices, and only two among the $\SU(N_S)$ indices, giving the following ten operators
\begin{align} \label{eq:drhosun}
O_{yD} ^{1-2} =& \frac{1}{16\pi^2} \left(y_{L} y_L^* \right)^{a_1} \phantom{} _{a_2}\phantom{} _{i_1} \phantom{}^{i_2} \left(y_{R}^* y_R \right)^{a_3} \phantom{} _{a_4}\phantom{} _{i_3} \phantom{}^{i_4}   \left(\Sigma \overleftrightarrow D ^\mu \Sigma^\dagger \right)_{a_1}\,^{a_2} \left(\Sigma^\dagger \overleftrightarrow D ^\mu \Sigma \right)_{a_3}\phantom{}^{a_4} \delta^{i_2}\, _{i_1} \delta^{i_4}\,_{i_3} \\
O_{yD} ^{3-4} =& \frac{1}{16\pi^2}\left(y_{L} y_L^* \right)^{a_1} \phantom{} _{a_2}\phantom{} _{i_1} \phantom{}^{i_2} \left(y_{R}^* y_R \right)^{a_3} \phantom{} _{a_4}\phantom{} _{i_3} \phantom{}^{i_4}   \left(\Sigma \overleftrightarrow D ^\mu \Sigma^\dagger \right)_{a_1}\,^{a_2} \left(\Sigma^\dagger \overleftrightarrow D ^\mu \Sigma \right)_{a_3}\phantom{}^{a_4} \delta^{i_2}\, _{i_1} \delta^{i_4}\,_{i_3} \\
O_{yD} ^{5-6} =& \frac{1}{16\pi^2}\left(y_{R}^* y_R \right)^{a_1} \phantom{} _{a_2}\phantom{} _{i_1} \phantom{}^{i_2} \left(y_{R}^* y_R \right)^{a_3} \phantom{} _{a_4}\phantom{} _{i_3} \phantom{}^{i_4}   \left(\Sigma ^\dagger\overleftrightarrow D ^\mu \Sigma \right)_{a_1}\,^{a_2} \left(\Sigma^\dagger \overleftrightarrow D ^\mu \Sigma \right)_{a_3}\phantom{}^{a_4} \delta^{i_2}\, _{i_1} \delta^{i_4}\,_{i_3} \\
 O_{yD} ^{7-8} = & \ \frac{1}{16\pi^2}\left(y_L^* y_L\right)_{a_1} \phantom{} ^{a_2 i_1} \phantom{} _{i_2}
 \left(y_R^* y_R\right)^{a_3} \phantom{} _{a_4 i_3} \phantom{} ^{i_4}\left(D_\mu \Sigma ^\dagger \right) _{a_3}\phantom{} ^{a_1} \left(D_\mu \Sigma   \right) _{a_2}\phantom{} ^{a_4} \delta^{i_2}\, _{i_1} \delta^{i_4}\,_{i_3} \\
 O_{yD}^{9-10} = & \ \frac{1}{16\pi^2} \left(y_L^* y_L\right)_{a_2} \phantom{} ^{a_3 i_2} \phantom{} _{i_3}  \left(y_R^* y_R\right)^{a_1} \phantom{} _{a_4 i_1} \phantom{} ^{i_4} \Sigma^\dagger _{a_1}\phantom{} ^{a_2}  \left(D_\mu \Sigma\right) _{a_3}\phantom{} ^{a_5}\left(\Sigma^\dagger D^\mu \Sigma\right)_{a_5} \, ^{a_4} \delta^{i_1}\, _{i_2} \delta^{i_3}\,_{i_4}
 \end{align}
%
Also in this case,  corrections stemming from the propagation of one SM gauge boson have to be considered. The latter are described by the same operators as those defined in eq.~\eqref{eq:NLOgauge}.

\bibliography{bibFPC}
\bibliographystyle{JHEP-2-2}

\end{document}